\documentclass[aps,prl,showpacs,twocolumn,superscriptaddress,floatfix]{revtex4-2}
\usepackage[utf8]{inputenc}
\newcommand{\figurescale}{1}
\usepackage{verbatim}
\usepackage{amsmath}
\usepackage{mathtools}

\DeclarePairedDelimiterX\braket[2]{\langle}{\rangle}{#1 \delimsize\vert #2}

\usepackage[pdftex]{graphicx}
\usepackage[separate-uncertainty=true]{siunitx}
\DeclareSIUnit{\rpm}{rpm}

\usepackage[colorlinks=true,urlcolor=blue,linkcolor=blue,citecolor=blue]{hyperref}
\usepackage[usenames,dvipsnames]{xcolor}
\usepackage[T1]{fontenc}
\usepackage{placeins}
\usepackage{nicefrac}
\usepackage{braket}
\usepackage{pdfcomment}
\usepackage{xcolor}
\usepackage{braket}
\usepackage[normalem]{ulem}

\usepackage{lineno}

\begin{document}

\title{Sensing the local magnetic environment through optically active defects in a layered magnetic semiconductor}
%
%
\author{J.~Klein}\email{jpklein@mit.edu}
\affiliation{Department of Materials Science and Engineering, Massachusetts Institute of Technology, Cambridge, MA 02139, USA}
\author{Z.~Song}
\affiliation{John A. Paulson School of Engineering and Applied Sciences, Harvard University, Cambridge, MA, USA}
\affiliation{College of Letters and Sciences, UCLA, Los Angeles, CA 90095 USA}
\author{B.~Pingault}
\affiliation{John A. Paulson School of Engineering and Applied Sciences, Harvard University, Cambridge, MA, USA}
\affiliation{QuTech, Delft University of Technology, 2600 GA Delft, The Netherlands}
\author{F.~Dirnberger}
\affiliation{Department of Physics, City College of New York, New York, NY 10031, USA}
\author{H.~Chi}
\affiliation{Francis Bitter Magnet Laboratory, Plasma Science and Fusion Center, Massachusetts Institute of Technology, Cambridge, MA 02139, USA}
\affiliation{U.S. Army CCDC Army Research Laboratory, Adelphi, Maryland 20783, USA}
\author{J.~B.~Curtis}
\affiliation{John A. Paulson School of Engineering and Applied Sciences, Harvard University, Cambridge, MA, USA}
\affiliation{College of Letters and Sciences, UCLA, Los Angeles, CA 90095 USA}
\author{R.~Dana}
\affiliation{Department of Materials Science and Engineering, Massachusetts Institute of Technology, Cambridge, MA 02139, USA}
\author{R.~Bushati}
\affiliation{Department of Physics, City College of New York, New York, NY 10031, USA}
\affiliation{Department of Physics, The Graduate Center, City University of New York, New York, NY 10016, USA}
\author{J.~Quan}
\affiliation{Department of Electrical and Computer Engineering, The University of Texas at Austin, Austin, TX, 78712, USA}
\affiliation{Photonics Initiative, CUNY Advanced Science Research Center, New York, NY, 10031, USA}
\affiliation{Department of Electrical Engineering, City College of the City University of New York, New York, NY, 10031, USA}
\affiliation{Physics Program, Graduate Center, City University of New York, New York, NY, 10026, USA}
\author{L.~Dekanovsky}
\affiliation{Department of Inorganic Chemistry, University of Chemistry and Technology Prague, Technická 5, 166 28 Prague 6, Czech Republic}
\author{Z.~Sofer}
\affiliation{Department of Inorganic Chemistry, University of Chemistry and Technology Prague, Technická 5, 166 28 Prague 6, Czech Republic}
\author{A.~Al\`{u}}
\affiliation{Department of Electrical and Computer Engineering, The University of Texas at Austin, Austin, TX, 78712, USA}
\affiliation{Photonics Initiative, CUNY Advanced Science Research Center, New York, NY, 10031, USA}
\affiliation{Department of Electrical Engineering, City College of the City University of New York, New York, NY, 10031, USA}
\affiliation{Physics Program, Graduate Center, City University of New York, New York, NY, 10026, USA}
\author{V.~M.~Menon}
\affiliation{Department of Physics, City College of New York, New York, NY 10031, USA}
\affiliation{Department of Physics, The Graduate Center, City University of New York, New York, NY 10016, USA}
\author{J.~S.~Moodera}
\affiliation{Francis Bitter Magnet Laboratory, Plasma Science and Fusion Center, Massachusetts Institute of Technology, Cambridge, MA 02139, USA}
\affiliation{Department of Physics, Massachusetts Institute of Technology, Cambridge, MA 02139, USA}
\author{M.~Lon\v{c}ar}
\affiliation{John A. Paulson School of Engineering and Applied Sciences, Harvard University, Cambridge, MA, USA}
\author{P.~Narang}\email{prineha@seas.harvard.edu}
\affiliation{John A. Paulson School of Engineering and Applied Sciences, Harvard University, Cambridge, MA, USA}
\affiliation{College of Letters and Sciences, UCLA, Los Angeles, CA 90095 USA}
\author{F.~M.~Ross}\email{fmross@mit.edu}
\affiliation{Department of Materials Science and Engineering, Massachusetts Institute of Technology, Cambridge, MA 02139, USA}
%
%
%
\date{\today}
%
%
\begin{abstract}
Atomic-level defects in van der Waals (vdW) materials are essential building blocks for quantum technologies and quantum sensing applications. The layered magnetic semiconductor CrSBr is an outstanding candidate for exploring optically active defects owing to a direct gap in addition to a rich magnetic phase diagram including a recently hypothesized defect-induced magnetic order at low temperature. Here, we show optically active defects in CrSBr that are probes of the local magnetic environment. We observe spectrally narrow ($\SI{1}{\milli\electronvolt}$) defect emission in CrSBr that is correlated with both the bulk magnetic order and an additional low temperature defect-induced magnetic order. We elucidate the origin of this magnetic order in the context of local and non-local exchange coupling effects. Our work establishes vdW magnets like CrSBr as an exceptional platform to optically study defects that are correlated with the magnetic lattice. We anticipate that controlled defect creation allows for tailor-made complex magnetic textures and phases with the unique ingredient of direct optical access.
\end{abstract}
%
%
\maketitle
%
%

Spin defects in solids make up a vastly growing field, both fundamental and applied, and play an important role in emergent quantum technologies.~\cite{ Degen.2017,Casola.2018,Atatre.2018,Liu.2019,Bhaskar.2020,Wolfowicz.2021,Hermans.2021,Philbin.2021} While single isolated defects in a solid are practical for emission of non-classical light or applications in quantum sensing,~\cite{Balasubramanian.2008,Casola.2018} closely arranged spin defects provide new means to explore complex quantum many-body systems in the solid state,~\cite{Choi.2020} similar to trapped atomic or ionic systems.~\cite{Bloch.2008} Such systems are of key interest for quantum simulation of spin-Hamiltonians for exploring exotic properties in solid-state materials.~\cite{Choi.2019,HeadMarsden.2020,Randall.2021,Peng.2021,Kennes.2021} 

In terms of local creation of defects with functional properties, the class of two-dimensional (2D) materials offers advantages over conventional three-dimensional (3D) materials.~\cite{Gottscholl.2021} Several optically active defects in 2D materials have been identified~\cite{MichaelisdeVasconcellos.2022} and also deterministically positioned~\cite{Klein.2019,Fournier.2021} with the motivation to provide a platform for scalable single-photon sources.~\cite{Branny.2017,PalaciosBerraquero.2017,Klein.2021a} 2D magnets offer novel opportunities for using defects for designing artificial magnetic orders or magnetic quasiparticles like magnetic vortices (e.g. skyrmions) for applications in nano-spintronics and quantum memories.~\cite{Lu.2020} As of yet, optically active defects in 2D magnets are scarcely explored~\cite{Gu.2019} often due to material instability, the absence of a band gap or poor optical emission properties~\cite{Huang.2017,Zhang.2018} and therefore mostly limited to theoretical work.~\cite{Lu.2020}

The layered magnetic semiconductor CrSBr is a promising vdW material ideal for pursuing 2D magnetism~\cite{Katscher.1966,Gser.1990,Beck.1990} due to its good air stability, sizeable band gap $\sim \SI{1.5}{\electronvolt}$,~\cite{Telford.2020} high transition temperature ($\SI{132}{\kelvin}$) with A-type antiferromagnetic (AFM) order,~\cite{Telford.2020}, tightly bound magneto-excitons~\cite{Wilson.2021} and correlated magneto-transport.~\cite{Telford.2022} Moreover, the charge transport is highly anisotropic~\cite{Wu.2022} even in multilayer CrSBr flakes which origin is a strong one-dimensional (1D) electronic character due to an intricate combination of weak interlayer hybridization and strong intralayer anisotropic electronic bandstructure.~\cite{Klein.2022} This has the advantage that optically clean signatures and magnetic orderings that mimic mono- or bilayer CrSBr can be conveniently probed in disorder free high-quality multilayer crystals.~\cite{Klein.2022}

\begin{figure*}
	\scalebox{\figurescale}{\includegraphics[width=1\linewidth]{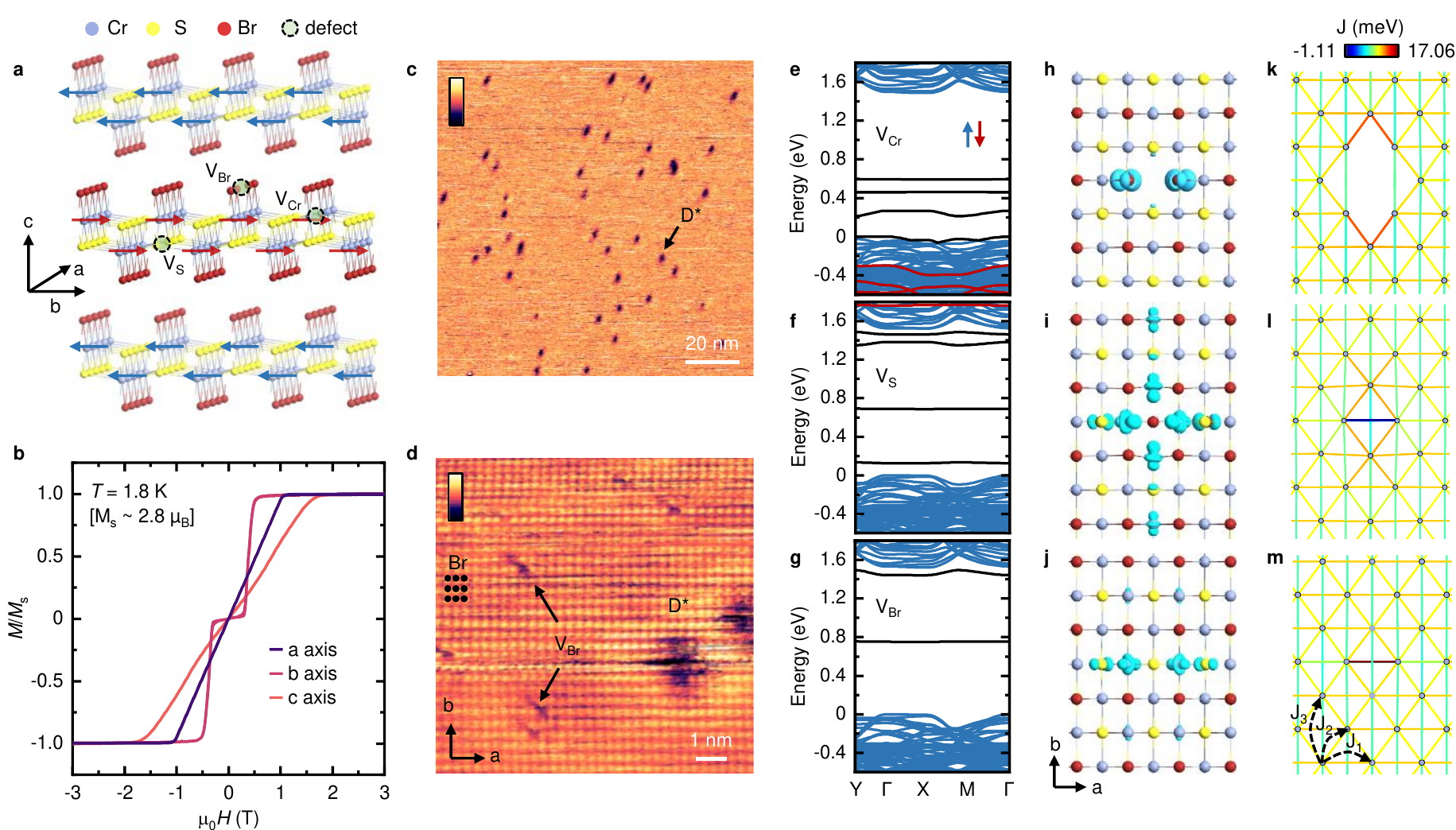}}
	\renewcommand{\figurename}{FIG.|}
	\caption{\label{fig1}
		\textbf{Intrinsic crystal defects in CrSBr and their calculated electronic and magnetic properties.}
		\textbf{a}, Schematic illustration of multilayer CrSBr with moment indicated in each layer and point vacancy defects within one layer.
		\textbf{b}, Magnetization at $\SI{1.8}{\kelvin}$ measured with the magnetic field oriented along the $a$, $b$ and $c$ axes. The magnetization saturates at $2.8 \mu_B$ close to the $S = + 3/2$ for ferromagnetic interlayer coupling.
		\textbf{c}, Large area room temperature STM topographic image of the surface defect concentration of at least one defect with a low density of $\sim 3 \cdot 10^{11}\SI{}{\per\centi\meter\squared}$. Tunneling current of $\SI{20}{\pico\ampere}$ and a bias voltage of $\SI{0.4}{\volt}$. 
		\textbf{d}, High magnification STM topographic image shows the top layer of Br atoms and individual Br vacancy point defect with a surface density of $\sim 5 \cdot 10^{12}\SI{}{\per\centi\meter\squared}$ (sheet density of $\sim 10^{13}\SI{}{\per\centi\meter\squared}$). 
		\textbf{e}, Ab initio calculated electronic band structure using the HSE functional of the V\textsubscript{Cr}, \textbf{f}, V\textsubscript{S} and \textbf{g} V\textsubscript{Br} for a 7 $\times$ 7 $\times$ 1 supercell.
		\textbf{h}, Top view of the calculated real-space wavefunctions of V\textsubscript{Cr}, \textbf{i} V\textsubscript{S} and \textbf{j} V\textsubscript{Br}.
		\textbf{k}, Calculated Heisenberg exchange interaction $J$ in presence of V\textsubscript{Cr}, \textbf{l} V\textsubscript{S} and \textbf{m} V\textsubscript{Br}.
		}
\end{figure*}

The magnetic phase diagram of CrSBr is rich, displaying a low temperature magnetic order that emerges at a critical temperature of $T_D = 30-\SI{40}{\kelvin}$.~\cite{Telford.2020,Telford.2022,Paz.2022,Boix-Constant.2022} This magnetic order has been observed consistently in magnetic susceptibility and magneto-transport measurements throughout all reported crystals. The origin of this magnetic signature is still under debate, but has been speculated to arise from crystal defects.~\cite{Telford.2022,Paz.2022,Boix-Constant.2022} The defect exchange coupling and defect-to-defect exchange interactions are therefore likely to play an important role in the formation of this magnetic order, but are as of yet unknown. Indeed, the type and density of intrinsic defects in CrSBr and their electronic, optical and magnetic properties are mostly unexplored. The strong optical response of this material,~\cite{Wilson.2021} in particular its very narrow spectral width in multilayer crystals,~\cite{Klein.2022} therefore creates a powerful motivation to study whether optically active defects are present in CrSBr, their sensitivity to the magnetic order and their role in the emergence of the low temperature magnetic order at $T_D = 30-\SI{40}{\kelvin}$. This will be crucial in exploiting crystal defects for applications in magnetic sensing and for creating exotic many-body states.


\begin{figure*}
	\scalebox{\figurescale}{\includegraphics[width=1\linewidth]{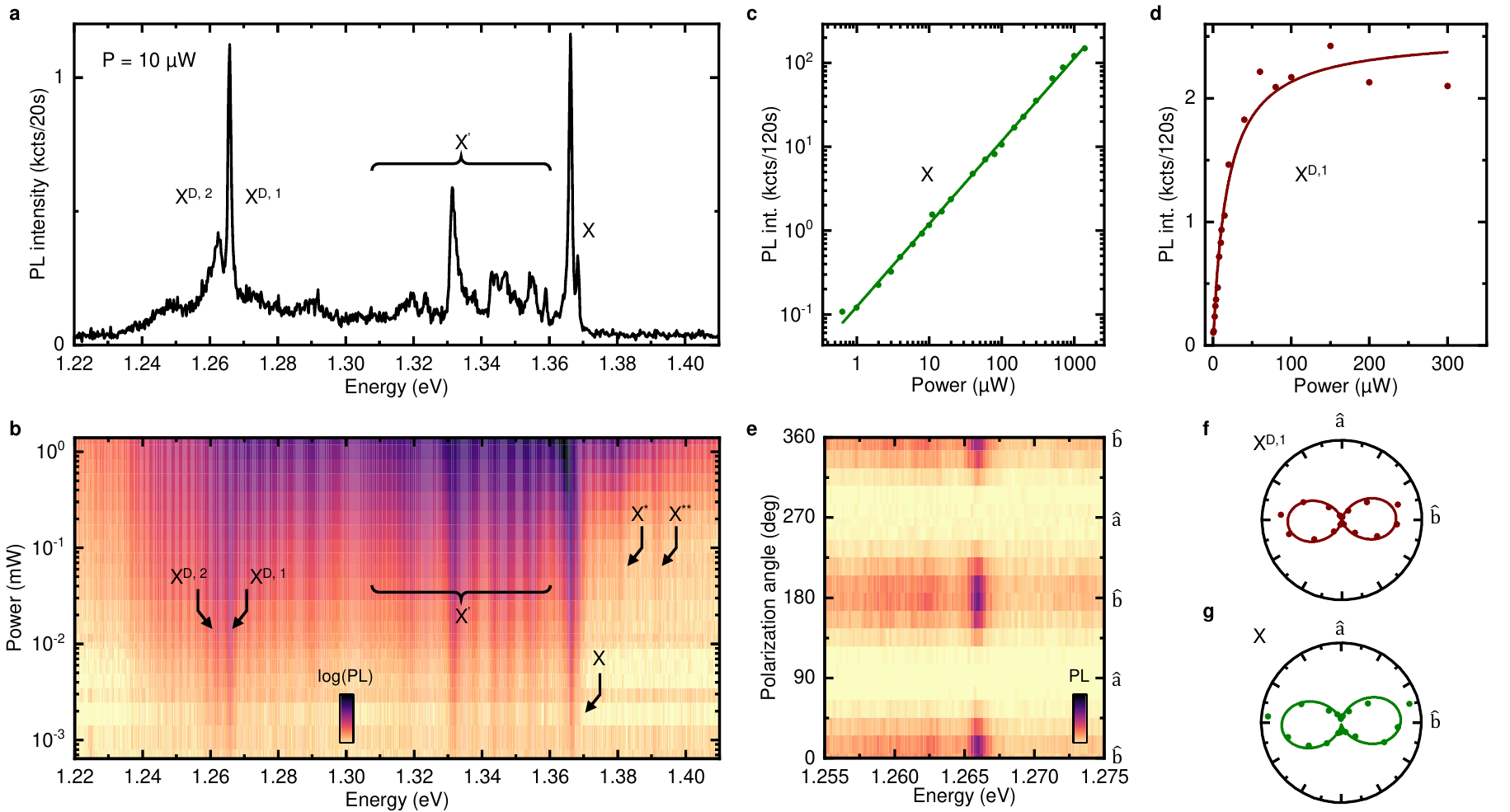}}
	\renewcommand{\figurename}{FIG.|}
	\caption{\label{fig2}
		\textbf{Spectrally narrow defect photoluminescence in CrSBr.}
		\textbf{a}, Low temperature ($\SI{4.2}{\kelvin}$) photoluminescence of multilayer CrSBr for an excitation power of $\SI{10}{\micro\watt}$ and a laser energy of $\SI{2.384}{\electronvolt}$. The 1s exciton $X$ and the defect doublet are highlighted.
		\textbf{b}, False color plot of the power dependence of the multilayer CrSBr spectrum.
		\textbf{c}, Linear power dependence of the 1s exciton ($X$).
		\textbf{d}, The defect peak shows a clear power saturation behavior with a saturation power of $P_S =17.99 \pm 2.15\SI{}{\micro\watt}$. This represents the saturation power for an ensemble of defects within the interaction volume of the laser spot.
		\textbf{e}, Polarization-dependent defect PL shows a strong linear polarization along the $b$ axis and the absence of PL along the $a$ axis. Corresponding polar plots for the defect emission ($X^{D,1}$) and exciton ($X$) emission are shown in \textbf{f} and \textbf{g}, respectively.
		}
\end{figure*}

Here, we show that CrSBr does indeed host optically active defect states that are correlated with the magnetic phase diagram. We quantify the intrinsic defects in CrSBr by scanning tunneling microscopy (STM) and observe that the most abundant defect is the V\textsubscript{Br} vacancy defect with a density of $\sim 10^{13}\SI{}{\per\centi\meter\squared}$ in addition to at least one more type of defect with a much lower density of $\sim 3 \cdot 10^{11}\SI{}{\per\centi\meter\squared}$ that is likely the V\textsubscript{S}. 
In low temperature photoluminescence (PL), we identify a spectral doublet that shows power saturation that we ascribe to emission from defects. This defect emission is magnetically correlated with the bulk magnetic order but shows up to $\sim 100$ times smaller energy changes as compared to the free excitons, reflecting the small but finite interlayer delocalization of the defect wavefunction when changing from an AFM to FM order. Most strikingly, we observe sharp crossover behaviour with the defect PL revealing a drastic spectral narrowing at $T_D$ that is associated with the emergence of the low temperature magnetic order. We calculate the electronic structure of the three most prolific point vacancy defects (V\textsubscript{Cr}, V\textsubscript{S} and V\textsubscript{Br}) and furthermore determine the impact of their presence on the local Heisenberg exchange coupling. Our experimental and theoretical results suggest that the low temperature magnetic order originates from the collective FM alignment of defects. The defects are either the origin of the phase, or `sense' this phase as optically active spin defects. The main mechanism creating this magnetic order is likely driven by a complex competition of thermal energy $k_B T$, local and non-local exchange interaction and charge carrier doping. These results overall suggest that strong opportunities exist for exploiting functional defects in CrSBr and vdW magnetic materials more generally, and provide a motivation for novel designs for the controlled generation of magnetic phases by atomic-level defect engineering in vdW magnets with the unique characteristic of direct optical detection. \\

\textbf{Intrinsic point vacancy defects in CrSBr.} We first discuss the nature of point defects in CrSBr, whose crystal structure and ground state magnetic ordering are schematically depicted in Fig.~\ref{fig1}a. CrSBr is an A-type antiferromagnet with an in-plane FM coupling and a weak interlayer AFM coupling. Measuring the magnetization of bulk CrSBr (see Fig.~\ref{fig1}b) shows the easy axis pointing along the $b$ axis with a spin flip transition at $B_{flip} = \SI{0.35}{\tesla}$, an intermediate magnetic axis along the $a$ axis ($\SI{1}{\tesla}$) and the hard axis along the $c$ axis ($\SI{2}{\tesla}$).~\cite{Gser.1990,Telford.2020} The three most prolific point vacancy defects are V\textsubscript{Cr}, V\textsubscript{S} and V\textsubscript{Br}, depicted in Fig.~\ref{fig1}a.

To measure the defect character and density we study a clean CrSBr surface in STM at room temperature. Figure~\ref{fig1}c shows a large area topographic image of CrSBr. We observe a low density ($\sim 3 \cdot 10^{11}\SI{}{\per\centi\meter\squared}$) of defects $D^*$ that have strong electronic contrast. 
These defects are not related to the surface but situated below the Br atoms, presumably within the Cr-S matrix, indicating that they originate from the V\textsubscript{S} or V\textsubscript{Cr} vacancy defect or a potentially more complex defect structure. Moreover, from high resolution topographic images (see Fig.~\ref{fig1}d) we determine a high top surface concentration of $\sim 5 \cdot 10^{12}\SI{}{\per\centi\meter\squared}$ of V\textsubscript{Br} that can be clearly distinguished by missing atoms in the periodically arranged Br atoms in the top surface. Their concentration corresponds to a sheet vacancy density of $\sim 10^{13}\SI{}{\per\centi\meter\squared}$, assuming that the top and bottom Br plane host the same number of V\textsubscript{Br}. This V\textsubscript{Br} concentration is almost two orders of magnitude higher than the other prevalent defects. The abundance of V\textsubscript{Br} is also thermodynamically expected from our calculated defect formation energies where $E^{Br}_{form} = \SI{3}{\electronvolt}$ is much smaller than $E^{S}_{form} = \SI{5}{\electronvolt}$ and $E^{Cr}_{form} = \SI{7}{\electronvolt}$, respectively. The high density of V\textsubscript{Br} can potentially explain the commonly observed n-type doping of CrSBr in magneto-transport~\cite{Telford.2020,Telford.2022,Wu.2022} with high sheet densities of $\sim 7 \cdot 10^{13}\SI{}{\per\centi\meter\squared}$.~\cite{Telford.2022} 

To study the effect of each vacancy on the local electronic environment, we performed ab initio calculations of a monolayer CrSBr with the Heyd-Scuseria-Ernzerhof hybrid (HSE) functional for the three most common defects V\textsubscript{Cr}, V\textsubscript{S} and V\textsubscript{Br} in a 7 $\times$ 7 $\times$ 1 supercell corresponding to a defect density of $\sim 1.2 \cdot 10^{13}\SI{}{\per\centi\meter\squared}$ (see Fig.~\ref{fig1}e-g). 
Based on these calculations, we expect that the V\textsubscript{Cr}, V\textsubscript{S} can induce four midgap states; we expect only two midgap states for V\textsubscript{Br}. The corresponding electronic wavefunctions of the V\textsubscript{Cr}, V\textsubscript{S} and V\textsubscript{Br} are shown in Fig.~\ref{fig1}h-i. The V\textsubscript{Cr} appears as the most localized in the $a$-$b$-plane and the V\textsubscript{S} the most delocalized along the $b$ axis, and the V\textsubscript{Br} the most delocalized along the $a$ axis with all defects following the D$_{2h}$ symmetry of CrSBr. We furthermore calculate the local change in the magnetic structure determining the local Heisenberg exchange interaction $J$ (see Fig.~\ref{fig1}k-m). The V\textsubscript{Cr} induces a magnetic hole in the lattice with slightly higher exchange energies as compared to the pristine lattice while the V\textsubscript{Br} shows the highest local exchange energy. We note for later reference that the local exchange of the V\textsubscript{S} is smaller and even negative (AFM) along the $a$ axis and that in particular the V\textsubscript{S} defect has a highly anisotropic wavefunction, with a large extent along the more dispersive $b$ axis.~\cite{Klein.2022}\\

\begin{figure}
	\scalebox{\figurescale}{\includegraphics[width=1\linewidth]{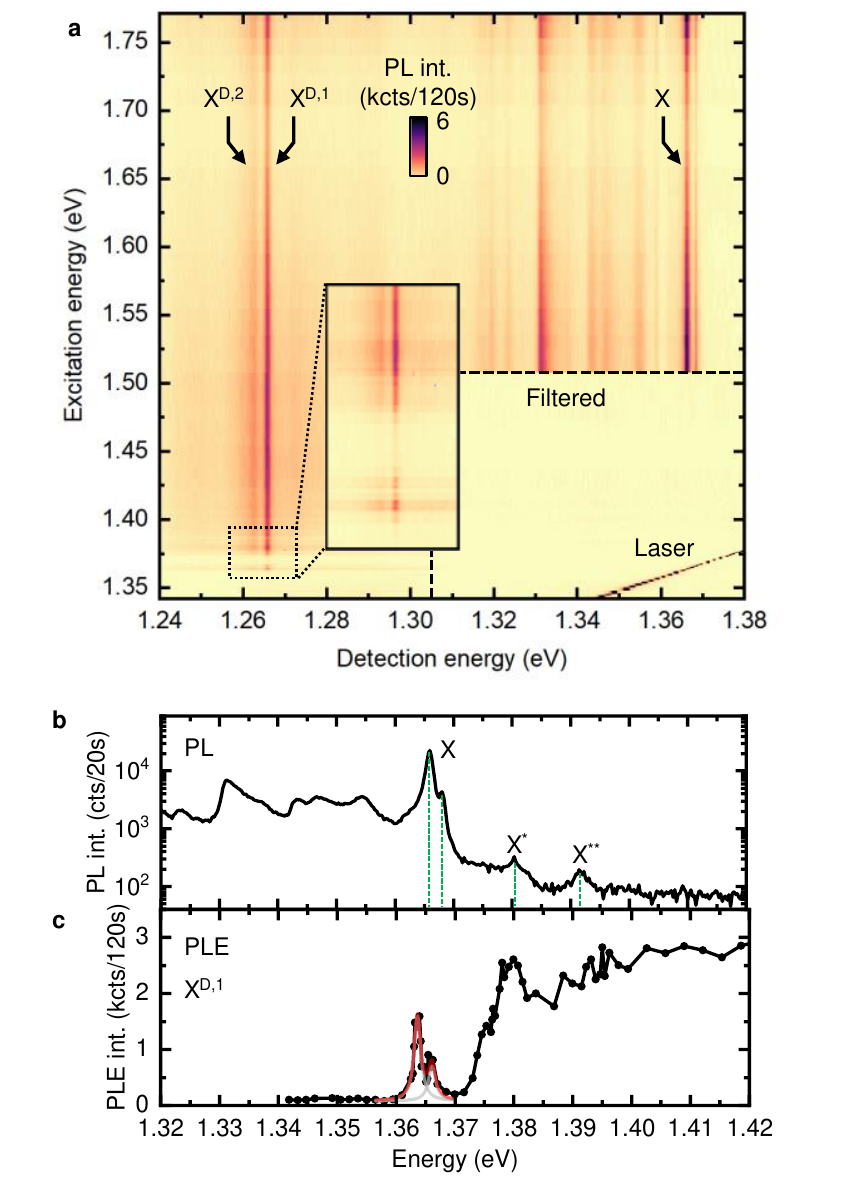}}
	\renewcommand{\figurename}{FIG.|}
	\caption{\label{fig3}
		\textbf{Electronic structure of defect emission in CrSBr.}
		\textbf{a}, Low temperature ($\SI{4.2}{\kelvin}$) PLE spectroscopy of multilayer CrSBr. The 1s exciton and defect doublet $X^{D}$ are indicated by arrows.
		\textbf{b}, PL spectrum on a semi-logarithmic scale highlighting excitonic transitions $X$, $X^{*}$ and $X^{**}$.
		\textbf{c}, Dependence of the defect intensity $X^{D,1}$ on the excitation energy. The resonances in the defect intensity reveal the excitonic structure of CrSBr. The red line is a fit to the excitonic doublet with a line width of $\SI{1.2}{\milli\electronvolt}$ and a splitting of $\SI{2.45}{\milli\electronvolt}$, in excellent agreement with the PL spectrum.
		}
\end{figure}


\textbf{Spectrally narrow defect emission in bulk CrSBr.} Having established the types and densities of defects present, we now measure their optical properties in multilayer CrSBr. For our PL measurements, we excite the material with a continuous-wave laser at an energy of $\SI{2.384}{\electronvolt}$ with the sample kept at a lattice temperature of $\SI{4.2}{\kelvin}$. A PL spectrum from a CrSBr flake with a thickness of $\SI{36.8}{\nano\meter}$ ($\sim 47$ layers) taken with an excitation power of $\SI{10}{\micro\watt}$ reveals a rich optical spectrum (see Fig.~\ref{fig2}a). The high crystal quality in combination with the preserved 1D character in the bulk provides very clean optical signatures with spectral linewidths of only $\SI{1}{\milli\electronvolt}$.~\cite{Klein.2022} From an excitation power dependence we can track the evolution of all emission lines (see Fig.~\ref{fig2}b). 

The spectrum shows emission from the 1s exciton ($X$) and additional resonances at higher energy ($X^{*}$ and $X^{**}$).~\cite{Klein.2022} The fine structure in $X$ is likely from interference effects due to the finite thickness of the flake.~\cite{Klein.2022} The sequence of optical features ($X^{\prime}$) in the energy window $\SI{50}{\milli\electronvolt}$ below $X$ is of unknown origin and not discussed further here.

While these resonances exhibit a linear power dependence (see e.g. $X$ in Fig.~\ref{fig2}c), the doublet peak labelled $X^{D,1}$ and $X^{D,2}$ at $\sim \SI{1.2658}{\electronvolt}$ and $\sim \SI{1.2624}{\electronvolt}$ shows a saturating power dependence (see Fig.~\ref{fig2}d). This is a clear distinction from free exciton emission and an unambiguous signature of emission from defects. The doublet fine structure with an energy splitting of $\SI{3.4}{\milli\electronvolt}$ is layer-independent, unlike the fine structure in the $X$, and therefore does not result from interference effects but is of real electronic origin (see SI). 



The excitation power dependent emission intensity is well described with $I(X^D) \sim \frac{A \cdot P}{P+P_{sat}}$ (see Fig.~\ref{fig2}d). We note that the emission of $X^D$ is not an individual defect but the emission from an ensemble of defects within the laser spot. Moreover, the defect emission is also strongly linearly polarized along the $b$ axis (see Fig.~\ref{fig2}e and f), similarly to the excitonic emission (see Fig.~\ref{fig2}g), indicating the anisotropic electronic structure. The calculated absorption of the V\textsubscript{S} is along the $b$ axis in closest agreement with the measured polarization (see SI). 

\begin{figure*}
	\scalebox{\figurescale}{\includegraphics[width=0.75\linewidth]{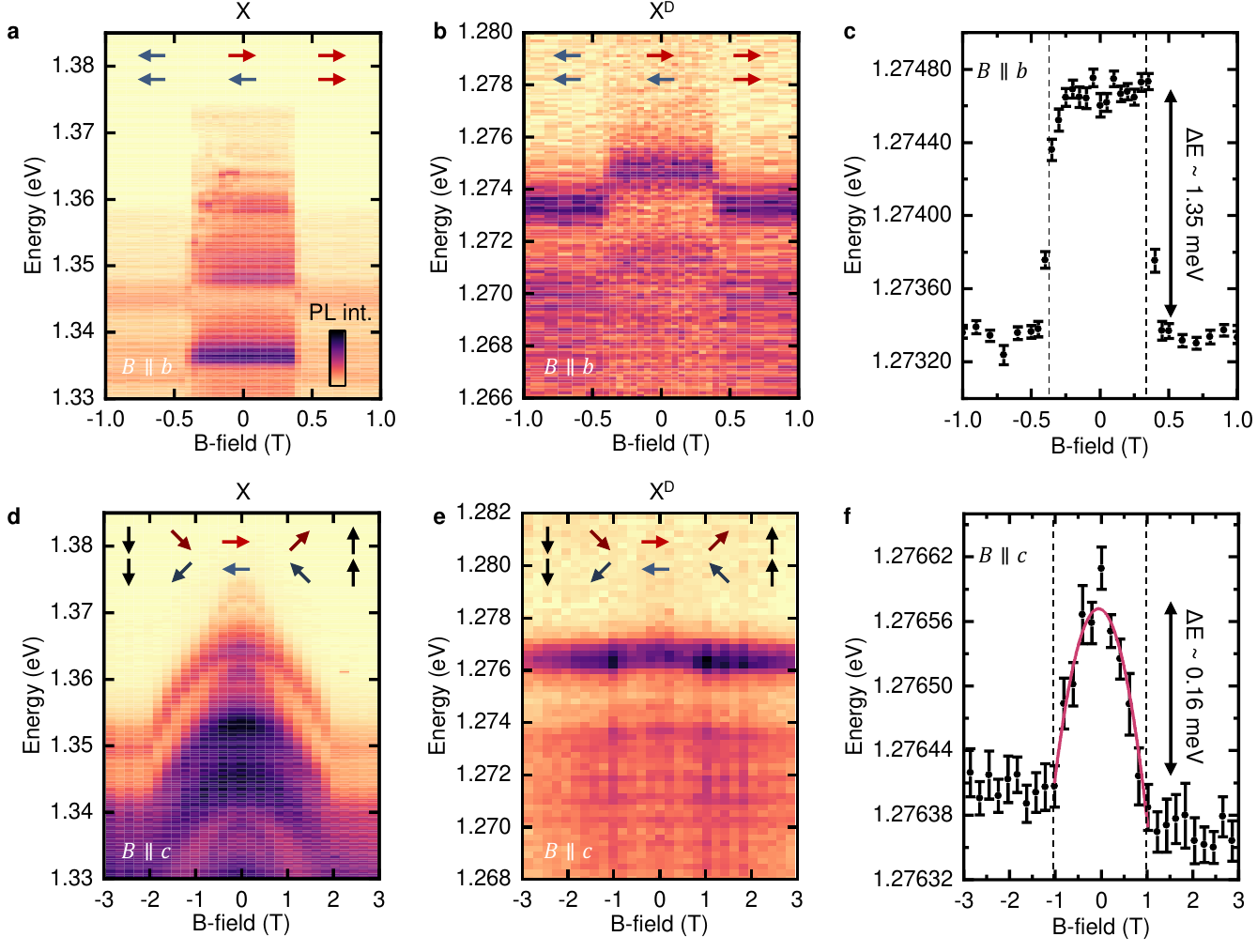}}
	\renewcommand{\figurename}{FIG.|}
	\caption{\label{fig4}
		\textbf{Sensing the magnetic order through optically active defects.}
		\textbf{a}, Magnetic field dependent PL of the excitons in a multilayer CrSBr for the magnetic field $B$ applied parallel to the $b$ axis. The exciton PL shifts when reaching the coercive field at $B_{flip} \sim \SI{0.35}{\tesla}$ due to a spin flip process where the AFM order changes to an FM order. Inset: AFM and FM ordering depicted in the reduced bilayer picture.
		\textbf{b}, Magnetic field dependent PL of the defect emission $X^D$ for the magnetic field $B$ applied parallel to the $b$ axis. The emission shows a sudden red-shift at the coercive field of the bulk CrSBr at $B_{flip} \sim \SI{0.35}{\tesla}$.
		\textbf{c}, Corresponding fitted emission energy of $X^{D,1}$ shows a small energy red-shift of $\SI{1.35}{\milli\electronvolt}$ at $B_{flip} \sim \SI{0.35}{\tesla}$
		\textbf{c}, Magnetic field dependent PL of the excitons for the magnetic field $B$ applied parallel to the $c$ axis. The exciton PL continuously shifts until reaching the coercive field at $B \sim \SI{2}{\tesla}$ due to spin canting along the $c$ axis. This changes from an in-plane AFM order to an out-of-plane FM order. Inset: AFM and FM ordering depicted in the reduced bilayer picture.
		\textbf{e}, Magnetic field dependent PL of the defect emission $X^D$ for the magnetic field $B$ applied parallel to the $c$ axis. The emission continuously red-shifts until a field of $B \sim \SI{1}{\tesla}$.
	    \textbf{f}, Corresponding emission energy of $X^{D,1}$ shows a small maximum energy red-shift of $\sim\SI{0.16}{\milli\electronvolt}$ at $B \sim \SI{1}{\tesla}$. The colored line is a parabolic fit to the data.
		}
\end{figure*}
%

\textbf{Electronic structure of the luminescent defect.} The relaxation dynamics and carrier pathways for different excitation energies can provide additional insights into the electronic structure of the luminescent defect. We probe the electronic structure of the defect in photoluminescence excitation (PLE) spectroscopy using a spectrally narrow ($\SI{1}{\nano\electronvolt}$) energy tunable continuous-wave Ti:Sapph laser. In our experiment, we study excitation energies ranging from $\SI{1.34}{\electronvolt}$ to $\SI{1.77}{\electronvolt}$. Figure~\ref{fig3}a shows a false color map of the PLE measurement. The PL intensity of the $X$ and the $X^D$ shows a similar dependence between $\SI{1.5}{\electronvolt}$ and $\SI{1.77}{\electronvolt}$ (see also full energy range in the SI). The increasing PL intensity at $\SI{1.77}{\electronvolt}$ is due to higher energy bands~\cite{Wilson.2021,Klein.2022} while the intensity increase at $\sim \SI{1.55}{\electronvolt}$ is likely from the high density of states (DOS) at the bulk single-particle band gap ($\sim 1.58 \SI{}{\electronvolt}$).~\cite{Klein.2022} More importantly, tuning the laser to the energy window where excitons dominate the material's response at $\SI{1.355}{\electronvolt}$ (see PL in Fig.~\ref{fig3}b) reveals a clear fine structure in the PLE of the defect emission $X^{D,1}$ (see Fig.~\ref{fig3}c and magnified inset in Fig.~\ref{fig3}a) consisting of a doublet at $\SI{1.3659}{\electronvolt}$ and $\SI{1.3684}{\electronvolt}$ and an additional peak at $\SI{1.381}{\electronvolt}$. Direct comparison with the PL spectrum obtained with an excitation energy of $\SI{1.7}{\electronvolt}$ at the same position is in excellent agreement with the $X$ doublet and the $X^{*}$. The third peak (labelled $X^{**}$) observed in PL is not fully resolved in the PLE, likely due to the smaller signal. The clearly resolved electronic and excitonic structure in the defect emission suggests that the defect can be excited efficiently by tuning the laser on resonance with points of high absorption in the electronic structure of CrSBr. The photo-excited electrons and holes can relax and populate the defect level efficiently followed by radiative recombination. \\

\textbf{Sensing the magnetic order through optically active defects.} It is now particularly interesting if the optically active defects embedded in the magnetic environment can be used as a probe of the local magnetic order. We therefore measure the defect photoluminescence as we control the magnetic order of the bulk crystal in an external magnetic field. Changes in the emission energy are expected to result from a change in magnetic order.~\cite{Wilson.2021} 

We begin by applying an external magnetic field along the $b$ axis (see Fig.~\ref{fig4}). The exciton shows the expected abrupt red-shift of $\sim \SI{20}{\milli\electronvolt}$ when the magnetic field exceeds the spin flip transition field of $B_{flip} = \SI{0.35}{\tesla}$ along the easy axis (see Fig.~\ref{fig4}a). This is in agreement with the sudden spin flip transition from AFM to FM order also observed in our magnetization measurements (see Fig.~\ref{fig1}b). Strikingly, the defect shows a qualitatively similar dependence with an energy red-shift at the same critical magnetic field (see Fig.~\ref{fig4}b). However, the magnitude of the energy shift is $\sim 15$ times lower than that of the exciton, only $\sim \SI{1.35}{\milli\electronvolt}$ (see Fig.~\ref{fig4}c). This small energy shift suggests that the wavefunction of the defect is significantly more localized within the layer but still exhibits a finite extent into the neighboring layer (i.e., a finite carrier tunneling rate $\tau$) when the magnetic order undergoes a transition from AFM to FM. 

We now apply the magnetic field along the $c$ axis (see Fig.~\ref{fig4}d). The exciton reveals a continuous red-shift due to the spins canting along the direction of the $B$-field until the saturation field of $\SI{2}{\tesla}$ is reached. In strong contrast, the defect emission exhibits a very small continuous energy red-shift (see Fig.~\ref{fig4}e and f) of only $\sim \SI{0.16}{\milli\electronvolt}$. The energy shift is a factor $\sim 100$ times smaller than that of the exciton. Moreover, the energy shift follows a parabolic dependence that agrees with the tunneling probability $\tau \propto B^2$ from second-order perturbation theory considerations.~\cite{Wilson.2021} Interestingly, the energy only shifts until reaching a magnetic field of $\SI{1}{\tesla}$, well below the saturation field of $\SI{2}{\tesla}$ for the exciton.

The magnitude of the shift is expected to correlate with the wavefunction delocalization in the FM order.~\cite{Wilson.2021} Generally, the sensitivity for energy shifts of the exciton in a magnetic field is due to the admixture of Cr $d$-orbitals with either in-plane or out-of-plane character.~\cite{Klein.2022} The observed difference in energy shift $\Delta E$ for the $b$ and $c$ axis of the defect emission is likely due to the different Cr $d$-orbital wavefunction admixture into the defect bands that are involved in the optical defect transition, and from the particular geometry of the defect wavefunction in real space (see Fig.~\ref{fig1}h-j). We calculate the orbital admixture into the defect bands of the V\textsubscript{Cr}, V\textsubscript{S} and V\textsubscript{Br} (see SI). The calculations suggest strong differences in $d$-orbital admixture for the three defects. In particular, the V\textsubscript{S} exhibits a strong admixture of Cr $d$-orbitals into its four defect bands. While all four defect bands share some out-of-plane character from $d_{z^2}$ orbitals, the two donor bands close to the conduction band have more in-plane character with $d_{(y^2-x^2)}$ while the two acceptor bands close to the valence band have out-of-plane character $d_{xz}$. The anisotropy in the energy shift points towards an anisotropy in the defect wavefunction in combination with the orbital admixture. In our case, the defect is more sensitive to out-of-plane changes than it is to in-plane changes in magnetic order. Such geometric considerations are of particular interest for the sensing of local magnetic fields. \\


\begin{figure*}
	\scalebox{\figurescale}{\includegraphics[width=1\linewidth]{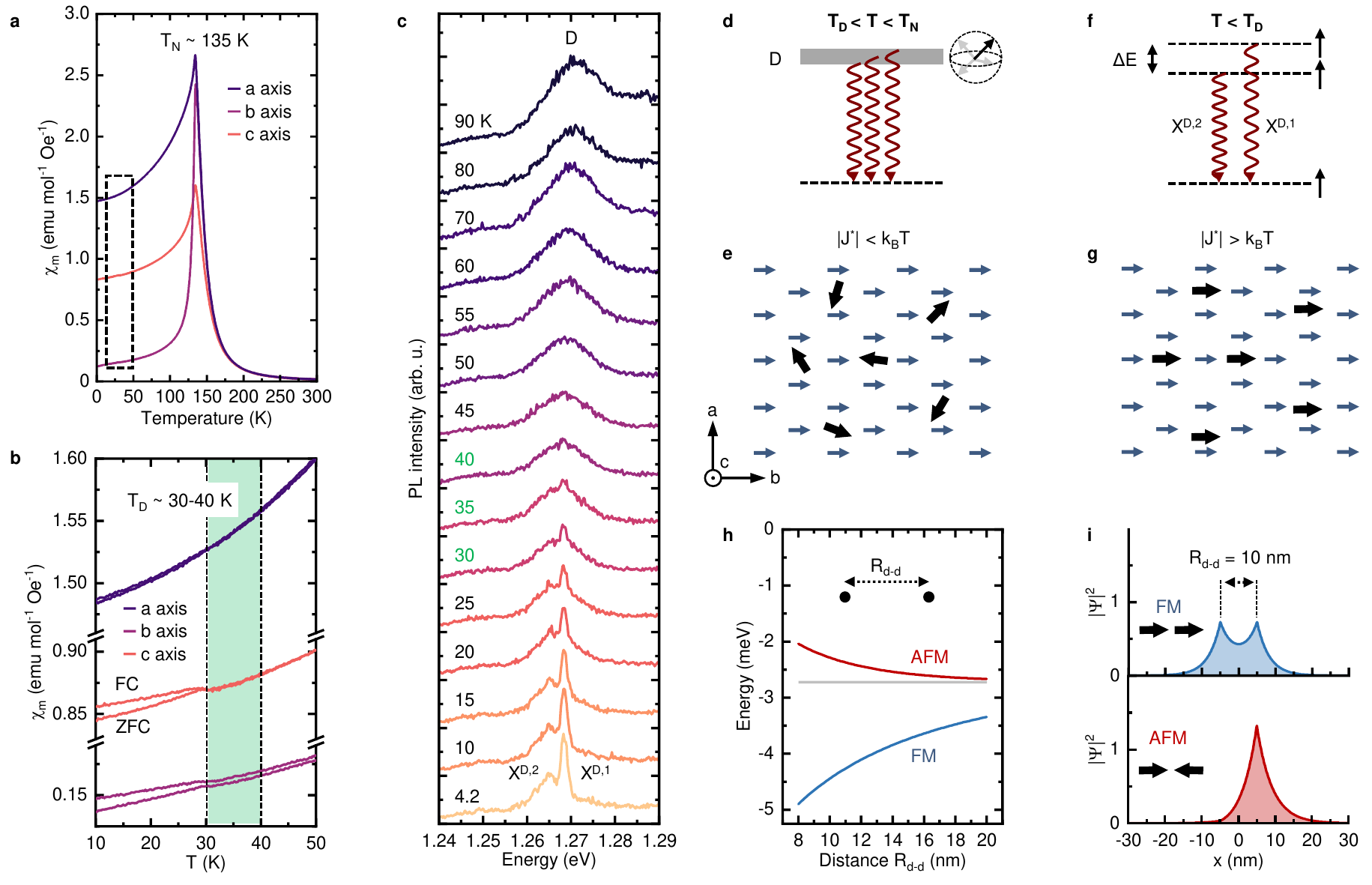}}
	\renewcommand{\figurename}{FIG.|}
	\caption{\label{fig5}
		\textbf{Low temperature defect magnetic order and optically active defects in CrSBr.}
		\textbf{a}, Magnetic susceptibility $\chi_m$ versus temperature along the $a$ axis, $b$ axis, and the $c$ axis for a CrSBr bulk crystal. Measurements are performed at a magnetic field of $\SI{10}{\milli\tesla}$.
		\textbf{b}, Zoom in from \textbf{c} showing a very weak signature of the low temperature magnetic order emerging at $30-40\SI{}{\kelvin}$ with a hysteresis behavior suggesting a FM order from field cooling (FC) and zero-field cooling (ZFC).
		\textbf{c}, Waterfall representation of the temperature evolution of the defect doublet PL.
		\textbf{d}, Schematic energy diagram of the uncoupled defect transition for $T_D < T < T_N$ exhibiting a random spin and as a result an energy broadened defect transition.
		\textbf{e}, Schematic illustration of the coupling mechanism of individual defect spins and the magnetic lattice. For a temperature $T_D < T < T_N$ the defect spin is not exchange coupled to the magnetic lattice because the defect-Cr exchange coupling $|J^{*}| < k_B T$.
		\textbf{f}, The FM alignment of the defect spin with the lattice manifests in a well-defined spin transition suggesting the narrowing of defect emission for $T < T_D$.
		\textbf{g}, For $T < T_D$ the defect spin is coupled via Heisenberg exchange to the surrounding Cr atoms due to $|J^{*}| > k_B T$. 
		\textbf{h}, Bound state energy for the two-spin aligned state (blue), the single spin aligned state (gray) and the spin anti-aligned state (red) as a function of distance between two defects $R_{d-d}$. The FM spin alignment is favorable.
		\textbf{i}, Polaron wavefunctions for different spin configurations for impurity separation $r = \SI{10}{\nano\meter}$ and scattering length $a = \SI{5}{\nano\meter}$. The FM spin wavefunction (blue) is lower in energy as compared to the AFM spin wavefunction (red) since it is able to delocalize more easily.
		}
\end{figure*}
%

\textbf{Low temperature magnetic order and optically active defect emitters.} After establishing that defects in CrSBr are local probes of their magnetic environment, we now focus on the role crystal defects play in the emergence of the defect-induced magnetic order in the temperature range $T_D = 30 - \SI{40}{\kelvin}$. We are particularly interested in how an ensemble of defects creates a magnetic order, how the defects interact with the magnetic lattice and also what role inter-defect interaction effects have in light of proximity effects and the experimentally measured defect densities. 


We begin with studying the correlation of the optically active defects with the low temperature magnetic order. Therefore, we measure the magnetic susceptibility $\chi_m$ by vibrating sample magnetometry (VSM) for field cooling (FC) and zero field cooling (ZFC) along the three main crystallographic axes (see Fig.~\ref{fig5}a and b). Our data clearly show the expected cusp of the Ne\'{e}l temperature at $T_N = \SI{135}{\kelvin}$. Moreover, we observe the magnetic order at the characteristic temperature $T_D = 30-40\SI{}{\kelvin}$, albeit with a very weak signal amplitude. This signature is universally observed in all crystals at around the same temperature.~\cite{Telford.2020,Telford.2022,Paz.2022,Boix-Constant.2022} The weaker signature in $\chi_m$ in combination with very narrow photoluminescence linewidths and a more intrinsic behavior from scanning tunneling spectroscopy~\cite{Klein.2022} together suggest a lower defect concentration in our CrSBr crystals.

We now examine the temperature dependence of the defect emission to track spectral changes in the temperature range of the low temperature magnetic order (see Fig.~\ref{fig5}c). Strikingly, upon warming from base temperature ($T = \SI{4.2}{\kelvin}$), the sharp spectral doublet of the defect $X^D$ disappears at $T_D = 30-40\SI{}{\kelvin}$, the temperature that is associated with the low temperature magnetic order. The simultaneous observation of $T_D$ where spectral narrowing occurs and the change in $\chi_m$ suggests that the emergence of the low temperature magnetic order is closely related to the optically active defects.


In the following we discuss the possible mechanism that can result in the defect-induced low temperature magnetic order. For our discussion we consider two interactions: (i) the direct exchange interaction of an isolated defect with the spins of the Cr lattice, and (ii) indirect interactions between defects that are mediated by free charge carriers or magnons.

We start with the magnetic order as a non-interacting magnetic state that emerges due to the individual alignment of defect spins below the critical temperature $T_D$. For any ordering to be established, the energy scale of the exchange interaction between two spins has to be larger than the thermal energy at a first approximation, disregarding other anisotropies. This is what usually defines critical transition temperatures in magnets. Similarly to this analogy, the energy scale of the exchange interaction $|J^{*}|$ between the defect spin and the surrounding spin lattice has to be larger than $k_B T$. Above the critical temperature but below the N\'{e}el temperature ($T_D < T < T_N$) the Cr spins are FM aligned in each layer. In this temperature window, since $k_B T_D > |J^{*}|$, the spin of the defects are subject to an exchange field but remain unpolarized, as thermal fluctuations overwhelm the exchange and lead to randomly oriented defect spins (see Fig.~\ref{fig5}d and e). However, when $k_B T_D < |J^{*}|$, the exchange interaction dominates over thermal disorder and can lead to a net polarization of the defect spins. The sign of $J^{*}$ then determines whether the defects align with an overall FM (positive) or AFM (negative) order. In our measurement, the alignment of defect spins in the critical temperature range $T_D \sim 30 - 40 \SI{}{\kelvin}$ is accompanied by a narrowing of the defect emission. This is likely due to inhomogeneous spin-dependent broadening of the defect energy, which changes from a broad distribution in the disordered phase (see Fig.~\ref{fig5}d and e) to a narrower distribution in the ordered phase (see Fig.~\ref{fig5}f and g). 

Given the strong and sharp crossover temperature in the defect emission (see Fig.~\ref{fig5}c) in combination with the hysteresis in $\chi_m$ (see Fig.~\ref{fig5}b), we suggest that the magnetic defect ordering is driven by additional defect-defect interactions. While the dominant effect is likely the exchange interaction between the defect spin state and the host magnetic lattice, we would expect this alone to result in a broad crossover behavior once $T\lesssim |J^*|$, reflecting independently distributed defects that remain uncorrelated. The observed sharp nonlinear behavior implies instead that defect-defect correlations also play a key role.



There are several plausible mechanisms that can introduce defect-defect interactions, including direct defect exchange, magnon-mediated defect interactions, or carrier-mediated RKKY-type interactions.~\cite{Kaminski.2002,Durst.2002,Calderon.2007,Priour.2006,Tang.2007,Liu.1986,Bednarski.2012} We specifically focus on the carrier-mediated origin here given the strong manifestation of the low temperature magnetic order in magneto-transport signatures,~\cite{Telford.2020,Telford.2022,Paz.2022,Boix-Constant.2022} but additional experiments are required to assess the applicability of the other scenarios, especially coupling to magnons.~\cite{Bae.2022}

To see how carriers can mediate the defect-defect interaction, we consider a simplistic 1D model featuring a single spin-polarized conduction band electron and two localized spin defects, separated by a distance $R_{d-d}$.~\cite{Kaminski.2002,Durst.2002,Liu.1986,Bednarski.2012} These defect spins then interact with the carrier by a local $s-d$ exchange interaction, with strength $K_{\rm d} < 0$ (the sign is unimportant except for determining the overall alignment of defect spins). This is plausible due to the 1D electronic character of CrSBr with its highly anisotropic conduction band and very light effective electron mass of $m^*_Y\sim 0.14 m_0$ in the dispersive $b$ axis~\cite{Klein.2022} and the dependence upon doping of the effect observed in magneto-transport.~\cite{Telford.2022}

We now use this model to compare the energies associated with the FM spin state and the AFM spin state as functions of the defect separation, as well as the single-defect binding energy (see Fig.~\ref{fig5}h and i). We determine the corresponding ground state energy of this complex considering the two spin values $S_1^z,S_2^z$ for each of the defects as a function of $R_{d-d}$ (see inset Fig.~\ref{fig5}h). In addition, we find that even for a single impurity, when the spin is aligned with the conduction band a bound state will form, and in the $\delta$-function limit this bound state energy is $E_1 = -\frac{\hbar^2}{8m^* a^2}$ where $a$ is the (bound-state) scattering length of the potential, determined by $1/a = m^* K_{\rm d}$ (see SI). The various binding energies for the bound states are shown in Fig.~\ref{fig5}h, for scattering length of $a = \SI{5}{\nano\meter}$. Indeed, we see a strong dependence on the distance between the two defects, and notably we find that while the AFM spin state tends towards essentially the single-defect energy, the FM spin state retains a noticeable lowering of the bound state energy compared to the single-defect energy, even at large distances. 

The interpretation is shown in Fig.~\ref{fig5}i, where we compute the corresponding wavefunctions for FM and AFM spin configurations for $R_{d-d} = \SI{10}{\nano\meter}$. Essentially, every aligned spin can lead to a bound state, but additionally when there are multiple aligned spins the bound states can hybridize more effectively by sharing their bound carriers, leading to a nonlinear contribution to the defect spin interactions. While this model offers a simple and compelling explanation for the origin of the low temperature magnetic order, it clearly requires a more detailed treatment, which we leave to future studies.

We can qualitatively compare the predictions of this model with the defect densities obtained from our topographic measurements (see Fig.~\ref{fig1}c and d) in which the inter-defect distances were $L_D = \frac{1}{\sqrt{\sigma}}\sim \SI{3}{\nano\meter}$ in case of V\textsubscript{Br} and a higher $L_D \sim \SI{18}{\nano\meter}$ for the other defects $D^{*}$. Both densities are within the range where sizeable interaction effects can be expected.

We finally discuss the possible origin of the defect emission and its relation to the low temperature magnetic order. Potential scenarios for defect emission include a defect-to-defect transition, conduction band-to-defect transition or defect-to-valence band transition. The defect emission $\sim\SI{100}{\milli\electronvolt}$ below $X$ suggests that the likely candidates are a defect-to-defect transition of the V\textsubscript{S} or a defect-to-valence band transition of V\textsubscript{Br} (see Fig.~\ref{fig1}f and g). The admixture of the valence band would result in more excitonic 'flavor' of the defect. Since we do not observe this in our measurements, this points more towards the V\textsubscript{S} that satisfies this requirement in our calculation with a defect level close to the conduction and valence band, respectively. This would also agree with the defect having less excitonic character with a more localized electron and hole wavefunction, also consistent with the magneto-PL and the temperature independent emission energy. Furthermore, the V\textsubscript{S} is in best agreement with the measured emission polarization along the $b$ axis of the defect emission as obtained from the calculated absorption (see SI). 



Moreover, the magnitude of the $\chi_m$ signal of the low temperature magnetic order suggests a high defect density. We observe a high concentration of V\textsubscript{Br}, with $\sim 1\%$ of missing atoms and a low concentration of defect $D^{*}$ with $\sim 0.05\%$ missing atoms, which likely is the V\textsubscript{S}. The underlying mechanism of the low temperature magnetic order can be more complex with more than one type of defect involved. We can imagine two potential scenarios: both the optical emission and the low temperature magnetic order arise exclusively from the V\textsubscript{Br}; or the magnetic order is from the V\textsubscript{Br} but sensed by the optically active V\textsubscript{S}. In general, the wavefunction anisotropy of V\textsubscript{S} along the 1D chains (dispersive $b$ axis) makes the coupling to free carriers that mediate exchange easier (see Fig.~\ref{fig1}i and l). While several experimental and theoretical results suggest that the most likely origin of optical defect emission is the V\textsubscript{S}, ascribing the exact origin of optically active defects is in general non-trivial. Additional experimental work is required to support its unambiguous identification. However, we note that all our results and interpretation are irrespective of the physical nature of the defect, showing only that different point defects can indeed have different roles. 



In conclusion, we experimentally show optically active defects in the CrSBr that sense the local magnetic environment, including a defect-induced magnetic order at low temperature. The optically active defect is either the origin or part of the mechanism that contributes to the low temperature magnetic order. Our results demonstrate that exciting new opportunities can emerge from the coupling of optically active defects with the underlying magnetic order of the matrix. Higher concentrations of defects can be interesting, if considered similarly to dopants or other stoichiometric changes. 

Atomic-level defects with the properties we have discussed here can serve as new means to engineer 2D magnetism and at the same time to probe the magnetic order via optical read-out. The optical signal and magnetic correlation has strong potential to harness the considerable advantages of optical experiments, such as probing non-equilibrium dynamics of magnetic systems in a simple experimental geometry. Finally, this work also motivates further efforts to explore the coupling of defects with magnetic quasiparticles like magnons~\cite{Bae.2022}


%
%
\section{Acknowledgements}
J.K. acknowledges support by the Alexander von Humboldt foundation. B.P. is a Marie Skłodowska-Curie fellow and acknowledges funding from the European Union’s Horizon 2020 research and innovation programme under the Grant Agreement No. 840968 (COHESiV). F.M.R. acknowledge the funding from the U.S. Department of Energy, Office of Basic Energy Sciences, Division of Materials Sciences and Engineering under Award DE‐SC0019336 for STEM characterization. Work by J.B.C. and P.N. is partially supported by the Quantum Science Center (QSC), a National Quantum Information Science Research Center of the U.S. Department of Energy (DOE). J.B.C. is an HQI Prize Postdoctoral Fellow and gratefully acknowledges support from the Harvard Quantum Initiative. Z.Song is supported through the Department of Energy BES QIS program on `Van der Waals Reprogrammable Quantum Simulator' under award number DE-SC0022277 for the work on long-range correlations, as well as partially supported by the Quantum Science Center (QSC), a National Quantum Information Science Research Center of the U.S. Department of Energy (DOE) on probing quantum matter. P.N. acknowledges support as a Moore Inventor Fellow through Grant No. GBMF8048 and gratefully acknowledges support from the Gordon and Betty Moore Foundation as well as support from a NSF CAREER Award under Grant No. NSF-ECCS-1944085. Z.Sofer was supported by project LTAUSA19034 from Ministry of Education Youth and Sports (MEYS) and by ERC-CZ program (project LL2101) from Ministry of Education Youth and Sports (MEYS).
L.D. was supported by specific university research (MSMT No. 20-SVV/2022). The VSM experiments were supported by Army Research Office (W911NF-20-2-0061 and DURIP W911NF-20-1-0074), National Science Foundation (NSF-DMR 1700137 and CIQM NSF-DMR 1231319) and Office of Naval Research (N00014-20-1-2306). H.C. was sponsored by the Army Research Laboratory under Cooperative Agreement Number W911NF-19-2-0015. The optical PL and PLE spectroscopy work at low temperatures was supported by ARO MURI (Grant No. W911NF1810432), HEADS-QON (Grant No. DE-SC0020376), ONR MURI (Grant No. N00014-15-1-2761), CIQM (Grant No. DMR-1231319), and ONR (Grant No. N00014-20-1-2425). Work at CCNY was supported through the NSF QII TAQS (V.M.M.) and the DARPA Nascent Light Matter program (R.B.). F.D. was funded by the Deutsche Forschungsgemeinschaft (DFG, German Research Foundation) through Projektnummer 451072703. A.A., J.Q. acknowledge support from a Vannevar Bush Faculty Fellowship, AFOSR DURIP and MURI programs and the Simons Foundation.

\section{Author contributions}
J.K. conceived the project and designed the experiments under supervision of F.M.R., J.K. prepared the samples, J.K. and B.P. performed PL and PLE optical measurements, F.D., R.B. and J.Q., performed magneto-PL measurements, H.C. and J.S.M. collected VSM data, R.D. and J.K. performed STM measurements, Z.Sofer and L.D. synthesized CrSBr crystals, J.K. analyzed the experimental data, Z.Song provided ab initio calculations, J.B.C. modelled the polaron exchange. Work by Z.Song and J.B.C. was supervised by P.N. V.M.M, A.A. and M.L. discussed results. J.K. wrote the manuscript with input from all co-authors.\\

%

\section{Methods}

\subsection{Sample fabrication}

CrSBr bulk crystals were grown by chemical vapor transport.~\cite{Klein.2021} Samples were fabricated by mechanical exfoliation onto SiO$_2$/Si substrates. Sample thickness was verified by atomic force microscopy, phase contrast and Raman spectroscopy for thinner flakes. 

\subsection{Photoluminescence spectroscopy}

For the PL measurements, we mounted the sample in closed-cycle helium cryostats (Montana Instruments or AttoDry 800) cryostat with a base temperature of $\SI{4.2}{\kelvin}$. In both setups, measurements were made through the side window using a home-built confocal microscope with a ×100, 0.9 NA objective (Olympus). We excited the sample using a continuous wave (CW)  laser at $\SI{2.384}{\electronvolt}$. The PL is collected confocally after using a long-pass filter. For the polarization resolved PL, we used a linear polarizer in both excitation and detection and a half-wave plate to rotate the polarization. 

\subsection{Photoluminescence excitation spectroscopy}

For the PLE measurements, we used a tunable CW Ti:Sapph laser (MSquared Solstis) with a linewidth of $~1\SI{}{\nano\electronvolt}$. The wavelength was stabilized using feedback from a wavemeter (High Finesse WS8). An excitation power of $\SI{10}{\micro\watt}$ was maintained throughout the measurement well below the saturation power of the defect emission.

\subsection{Magneto-optical spectroscopy}

Magnetic field-dependent optical measurements were conducted by mounting a sample of CrSBr bulk flakes on top of a standard $\textrm{SiO}_2/\textrm{Si}$ substrate into a closed-cycle cryostat (AttoDry 2100). The sample was subsequently cooled to temperatures around $\SI{1.6}{\kelvin}$. To align $b$ or $c$ crystal axes of individual bulk flakes along the axis of the superconducting solenoid magnet, providing field strengths up to $\SI{9}{\tesla}$, the sample was mounted onto a horizontal sample holder ($c$ axis), or onto a perpendicular sample holder ($b$ axis) and visually aligned. We used a $\SI{2.33}{\electronvolt}$ continuous-wave laser that was fiber-coupled on the input to perform magnetic field-dependent PL measurements. A second fiber collected the signal from the sample and directed it towards a high-resolution spectrometer attached to a liquid-Nitrogen cooled charge-coupled device camera. The laser excitation power was $\SI{100}{\micro\watt}$.

\subsection{Magnetization measurements}

Temperature, field and angle dependent magnetic measurements were performed in the temperature range of $2 - \SI{300}{K}$ in a Quantum Design Physical Property Measurement System (PPMS) equipped with a $\SI{9}{\tesla}$ superconducting magnet. 
Vibrating sample magnetometry (VSM) was employed to characterize the magnetization. 

\subsection{Scanning tunneling microscopy}

The topographic images were taken at room temperature with a Unisoku UHV-LT four-probe scanning tunnelling microscope operated with a Nanonis controller. The STM is equipped with a scanning electron microscope that allows precise location of scan locations. The CrSBr bulk crystal was cleaved in vacuum to obtain an adsorbate-free and clean surface, and the data were acquired with both PtIr and W tips. 

\subsection{Ab inito calculations}

The band structure calculations for the three vacancy defects were calculated using SG15 pseudopotential~\cite{Schlipf.2015} and the Heyd-Scuseria-Ernzerhof hybrid (HSE) exchange correlation functional~\cite{Heyd.2004}. An atomic basis set was used and spin-orbit coupling was turned on. We used a momentum space $k$-point sampling of 4 $\times$ 3 $\times$ 1 for our calculation and a supercell of 7$\times$ 7 $\times$ 1 to avoid interaction between defects.

The optical absorption was calculated using the Perdew–Burke-Ernzerhof (PBE) exchange correlation functional~\cite{Perdew.1996} and a supercell of 7$\times$ 7 $\times$ 1.

\subsection{Heisenberg exchange}

We investigated the ferromagnetism in monolayer CrSBr in presence of vacancy defects by calculating the isotropic Heisenberg exchange coupling constant of the nearest neighbor $J_{1,2,3}$ using the Liechtenstein method.~\cite{terasawa2019efficient} For the calculation we used a 5 $\times$ 5 $\times$ 1 supercell. For the pristine CrSBr without a defect we obtain exchange coupling constants of $J_1 = \SI{8.21}{\electronvolt}$, $J_2 = \SI{7.04}{\electronvolt}$ and $J_3 = \SI{2.28}{\electronvolt}$. In our notation a positive value represents FM coupling while a negative value corresponds to AFM coupling. Our calculated values are overestimated in comparison to other experimental and theoretical works~\cite{Wang.2020,Scheie.2022} but agree with the expected trend that $J_1$ $\sim$ $J_2$ > $J_3$.


%
%

\bibliographystyle{naturemag}
\bibliography{full}

\end{document}


\title{SI - Sensing the local magnetic environment through optically active defects in a layered magnetic semiconductor}
%
%
\author{J.~Klein}\email{jpklein@mit.edu}
\affiliation{Department of Materials Science and Engineering, Massachusetts Institute of Technology, Cambridge, MA 02139, USA}
\author{Z.~Song}
\affiliation{John A. Paulson School of Engineering and Applied Sciences, Harvard University, Cambridge, MA, USA}
\affiliation{College of Letters and Sciences, UCLA, Los Angeles, CA 90095 USA}
\author{B.~Pingault}
\affiliation{John A. Paulson School of Engineering and Applied Sciences, Harvard University, Cambridge, MA, USA}
\affiliation{QuTech, Delft University of Technology, 2600 GA Delft, The Netherlands}
\author{F.~Dirnberger}
\affiliation{Department of Physics, City College of New York, New York, NY 10031, USA}
\author{H.~Chi}
\affiliation{Francis Bitter Magnet Laboratory, Plasma Science and Fusion Center, Massachusetts Institute of Technology, Cambridge, MA 02139, USA}
\affiliation{U.S. Army CCDC Army Research Laboratory, Adelphi, Maryland 20783, USA}
\author{J.~B.~Curtis}
\affiliation{John A. Paulson School of Engineering and Applied Sciences, Harvard University, Cambridge, MA, USA}
\affiliation{College of Letters and Sciences, UCLA, Los Angeles, CA 90095 USA}
\author{R.~Dana}
\affiliation{Department of Materials Science and Engineering, Massachusetts Institute of Technology, Cambridge, MA 02139, USA}
\author{R.~Bushati}
\affiliation{Department of Physics, City College of New York, New York, NY 10031, USA}
\affiliation{Department of Physics, The Graduate Center, City University of New York, New York, NY 10016, USA}
\author{J.~Quan}
\affiliation{Department of Electrical and Computer Engineering, The University of Texas at Austin, Austin, TX, 78712, USA}
\affiliation{Photonics Initiative, CUNY Advanced Science Research Center, New York, NY, 10031, USA}
\affiliation{Department of Electrical Engineering, City College of the City University of New York, New York, NY, 10031, USA}
\affiliation{Physics Program, Graduate Center, City University of New York, New York, NY, 10026, USA}
\author{L.~Dekanovsky}
\affiliation{Department of Inorganic Chemistry, University of Chemistry and Technology Prague, Technická 5, 166 28 Prague 6, Czech Republic}
\author{Z.~Sofer}
\affiliation{Department of Inorganic Chemistry, University of Chemistry and Technology Prague, Technická 5, 166 28 Prague 6, Czech Republic}
\author{A.~Al\`{u}}
\affiliation{Department of Electrical and Computer Engineering, The University of Texas at Austin, Austin, TX, 78712, USA}
\affiliation{Photonics Initiative, CUNY Advanced Science Research Center, New York, NY, 10031, USA}
\affiliation{Department of Electrical Engineering, City College of the City University of New York, New York, NY, 10031, USA}
\affiliation{Physics Program, Graduate Center, City University of New York, New York, NY, 10026, USA}
\author{V.~M.~Menon}
\affiliation{Department of Physics, City College of New York, New York, NY 10031, USA}
\affiliation{Department of Physics, The Graduate Center, City University of New York, New York, NY 10016, USA}
\author{J.~S.~Moodera}
\affiliation{Francis Bitter Magnet Laboratory, Plasma Science and Fusion Center, Massachusetts Institute of Technology, Cambridge, MA 02139, USA}
\affiliation{Department of Physics, Massachusetts Institute of Technology, Cambridge, MA 02139, USA}
\author{M.~Lon\v{c}ar}
\affiliation{John A. Paulson School of Engineering and Applied Sciences, Harvard University, Cambridge, MA, USA}
\author{P.~Narang}\email{prineha@seas.harvard.edu}
\affiliation{John A. Paulson School of Engineering and Applied Sciences, Harvard University, Cambridge, MA, USA}
\affiliation{College of Letters and Sciences, UCLA, Los Angeles, CA 90095 USA}
\author{F.~M.~Ross}\email{fmross@mit.edu}
\affiliation{Department of Materials Science and Engineering, Massachusetts Institute of Technology, Cambridge, MA 02139, USA}
%
%
\date{\today}
%

%
\maketitle
%
%

\tableofcontents

\newpage


\section{Scanning tunneling microscopy of CrSBr}

%
\begin{figure*}[ht]
\scalebox{\figurescale}{\includegraphics[width=1\linewidth]{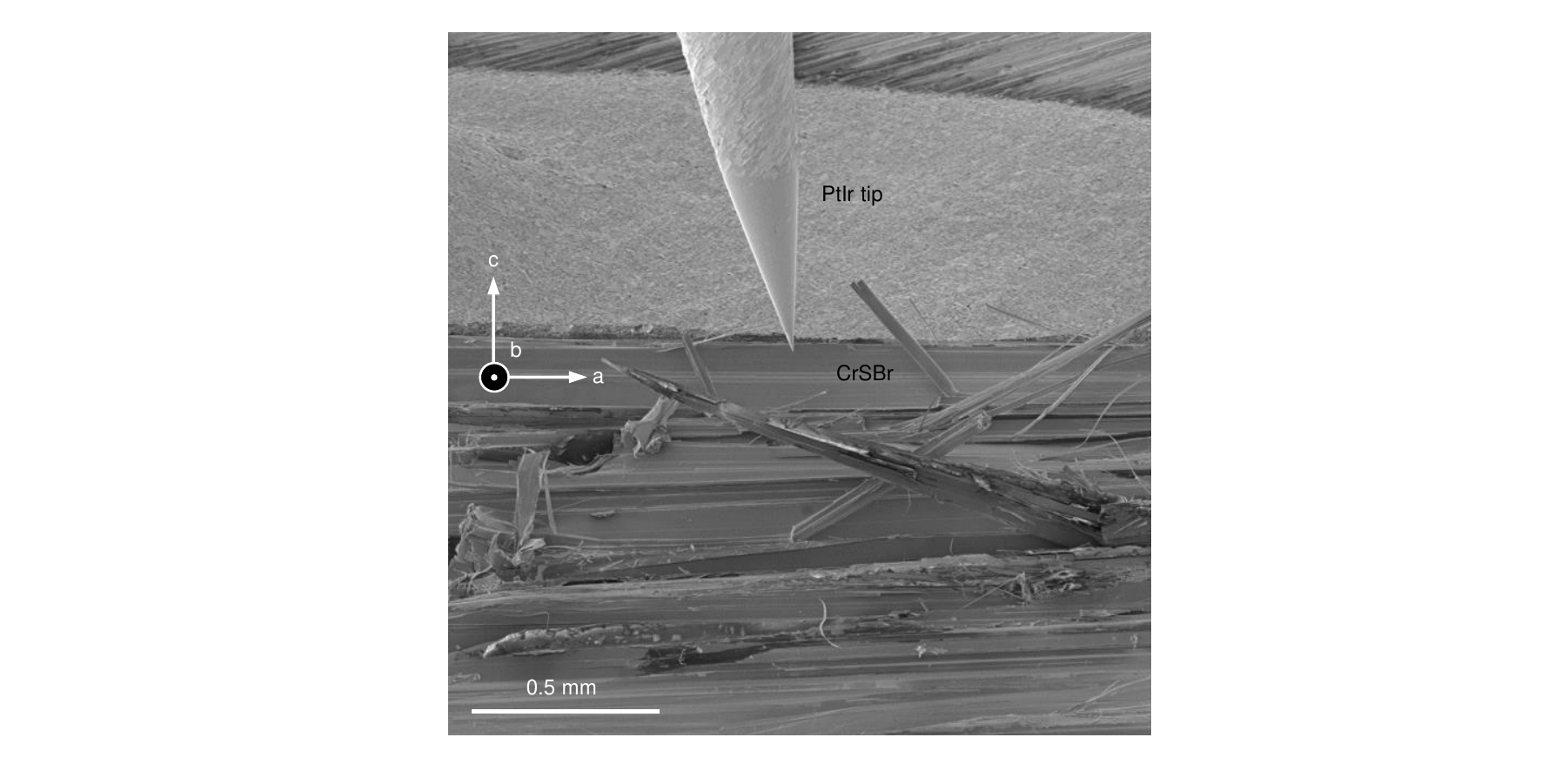}}
\renewcommand{\figurename}{SI Fig.|}
\caption{\label{SISTM}
\textbf{Scanning electron microscope (SEM) image of CrSBr in the STM.} 
Typical SEM image of a bulk CrSBr crystal that was exfoliated in situ in the STM under high vacuum. The CrSBr forms needle like flakes with the long axis along the $a$ axis and the short axis along the $b$ axis. We locate clean surfaces for landing the PtIr tip using the SEM.
}
\end{figure*}
%
\newpage

\section{Power dependent exciton and defect emission}

%
\begin{figure*}[ht]
\scalebox{\figurescale}{\includegraphics[width=1\linewidth]{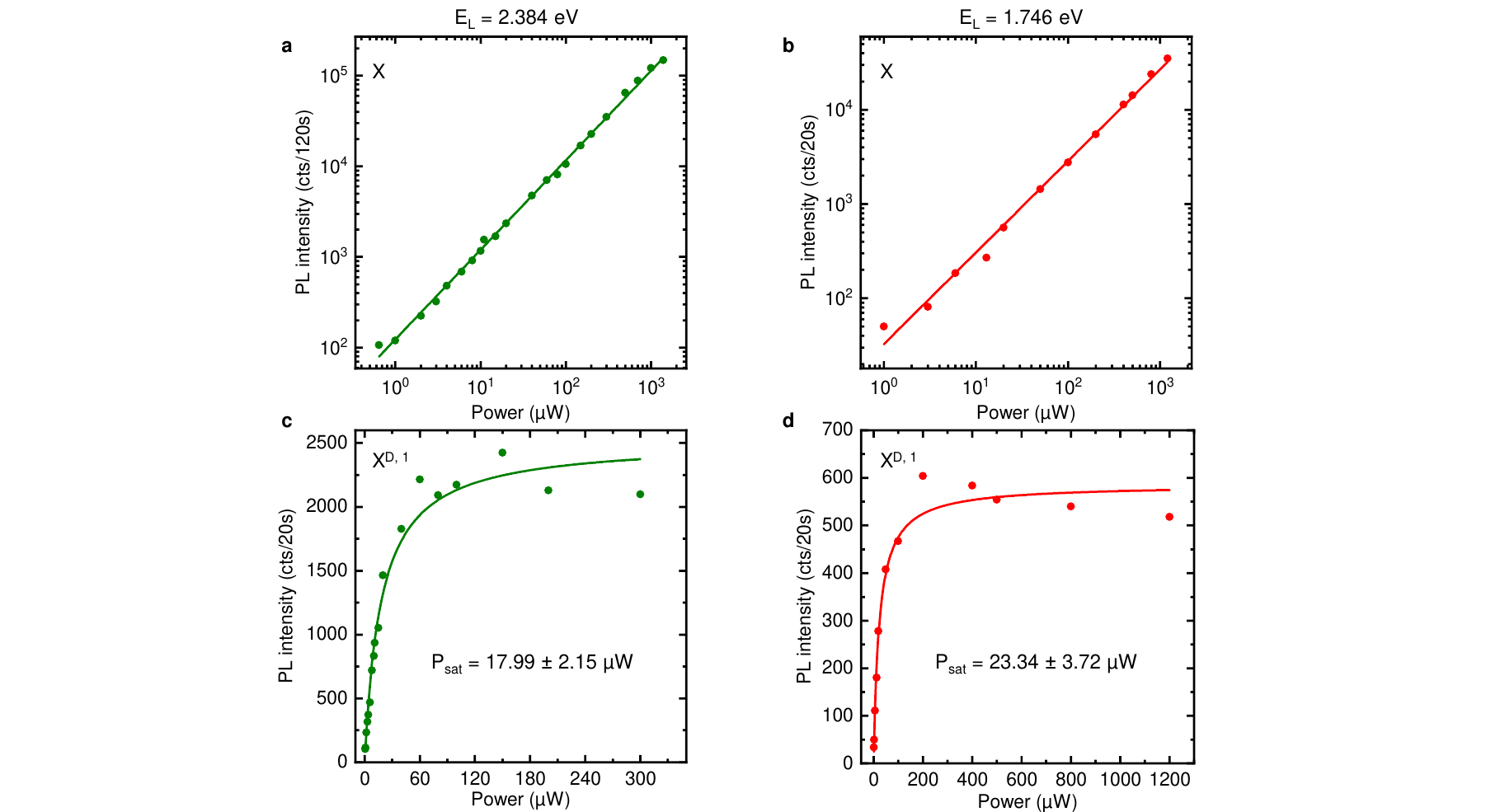}}
\renewcommand{\figurename}{SI Fig.|}
\caption{\label{SIPower}
\textbf{Power dependence of exciton and the defect emission.} 
\textbf{a}, Power dependence of the exciton on a double logarithmic plot for an excitation laser energy of $E_L = \SI{2.384}{\electronvolt}$ and \textbf{b}, an excitation laser energy of $E_L = \SI{1.746}{\electronvolt}$. The PL intensity follows a linear dependence with a slope of $0.99 \pm 0.01$ and $0.97 \pm 0.04$, respectively.
\textbf{c}, Power dependence of the defect emission $X^{D, 1}$ for an excitation laser energy of $E_L = \SI{2.384}{\electronvolt}$ and \textbf{d} an excitation laser energy of $E_L = \SI{1.746}{\electronvolt}$. The solid line is a fit with $I = A \cdot P / (P+P_{sat})$ resulting in a saturation power of $P_{sat} = 17.99 \pm 2.15 \SI{}{\micro\watt}$ and $P_{sat} = 23.34 \pm 3.72 \SI{}{\micro\watt}$, respectively.
}
\end{figure*}
%
\newpage

\section{Layer independent energy splitting of the defect emission}

%
\begin{figure*}[ht]
\scalebox{\figurescale}{\includegraphics[width=1\linewidth]{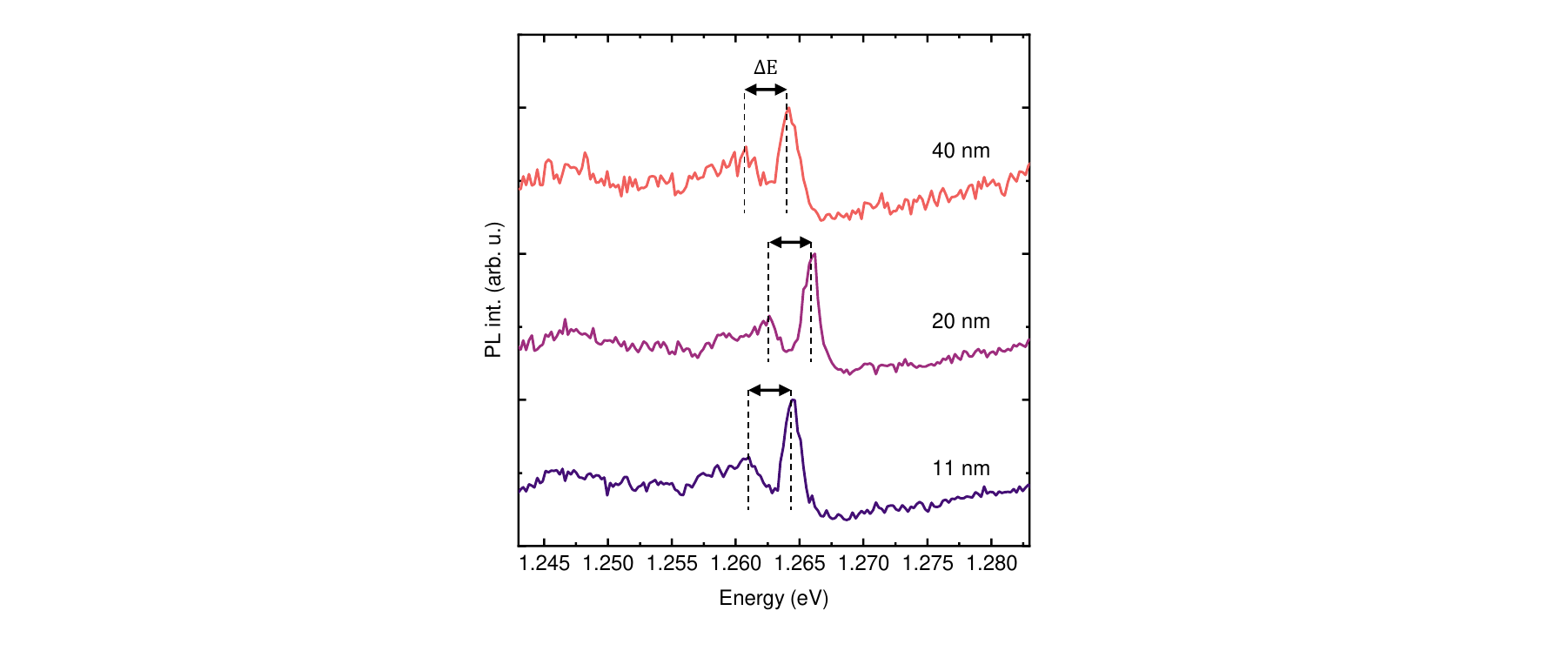}}
\renewcommand{\figurename}{SI Fig.|}
\caption{\label{SILayer}
\textbf{Thickness-independent defect emission.} 
Low temperature ($\SI{4.2}{\kelvin}$) PL spectra of CrSBr with thicknesses of $\SI{11}{\nano\meter}$, $\SI{20}{\nano\meter}$ and $\SI{40}{\nano\meter}$. Thicknesses are obtained by atomic force microscopy. The energy splitting of the defect emission is constant with an energy of $\Delta E = \SI{3.6}{\milli\electronvolt}$ and independent of the layer thickness.
}
\end{figure*}
%

\newpage

\section{Photoluminescence excitation spectroscopy and linewidth of defect emission}

%
\begin{figure*}[ht]
\scalebox{\figurescale}{\includegraphics[width=1\linewidth]{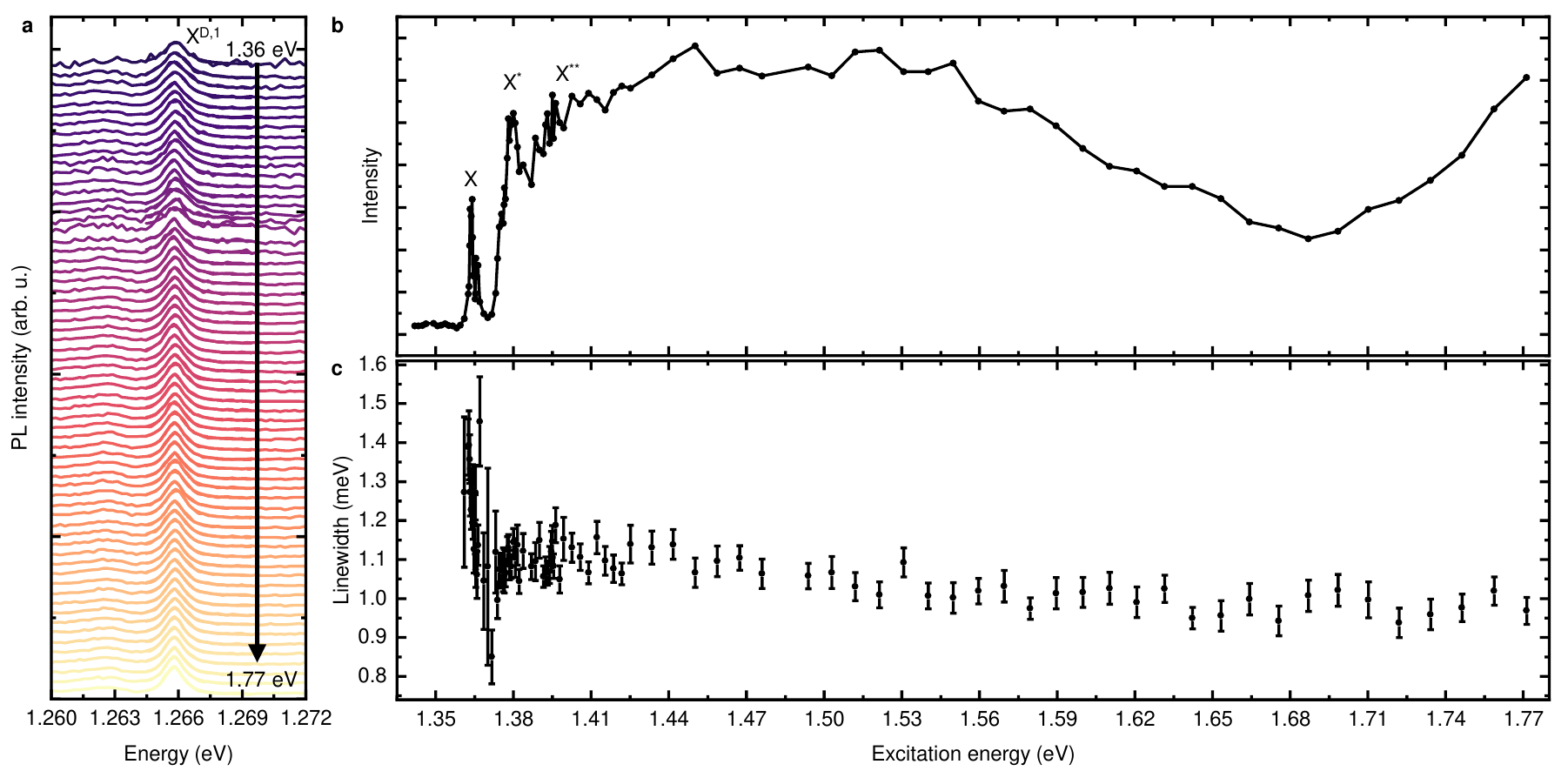}}
\renewcommand{\figurename}{SI Fig.|}
\caption{\label{SIfigPLE}
\textbf{Photoluminescence excitation spectroscopy of the defect emission in CrSBr.} 
\textbf{a}, Waterfall plot of the normalized PL of the defect emission. Data are collected for excitation energies ranging from $\SI{1.36}{\electronvolt}$ to $\SI{1.77}{\electronvolt}$. The solid line is a Lorentzian fit to the energetically higher lying peak of the defect doublet $X^{D,1}$.
\textbf{b}, PL intensity of the defect emission as a function of the excitation laser energy. The PLE scan reveals the exciton doublet and higher lying excitonic states and a high density of states up to an energy of $\SI{1.65}{\electronvolt}$. The increasing intensity towards higher energies is from higher lying bands that contribute to a high density of states.
\textbf{c}, Linewidth of the Lorentzian fits from \textbf{a} as a function of excitation laser energy. The linewidth at higher energies ($E > \SI{1.4}{\electronvolt}$) is $\sim \SI{1}{\milli\electronvolt}$ but monotonically increases to a maximum value of $\SI{1.5}{\milli\electronvolt}$ when the laser energy is tuned over the exciton doublet.
}
\end{figure*}
%
\newpage

\section{Temperature dependent energy shift}

%
\begin{figure*}[ht]
\scalebox{\figurescale}{\includegraphics[width=1\linewidth]{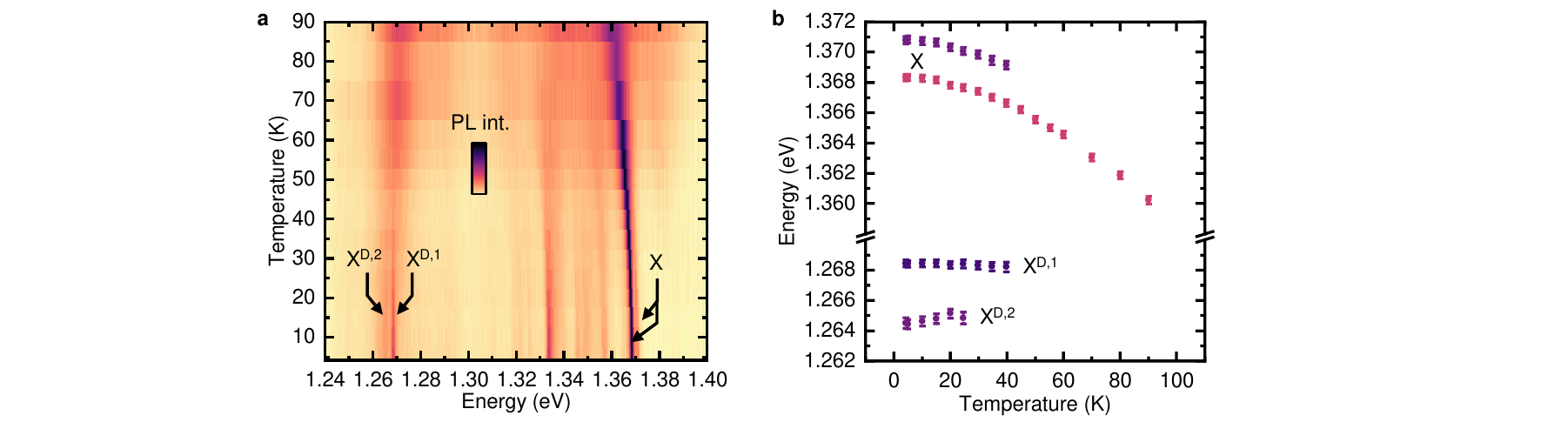}}
\renewcommand{\figurename}{SI Fig.|}
\caption{\label{SIfigTemp}
\textbf{Temperature evolution of the exciton and defect emission.} 
\textbf{a}, False color plot of the temperature-dependent PL of multilayer CrSBr.
\textbf{b}, Fitted emission energy of the 1s exciton doublet (from interference effect) and the defect doublet (real electronic effect) as a function of temperature. The exciton shows a clear Varshni behaviour with an energy red-shift for increasing temperature. In contrast, the defect doublet exhibits a constant energy with changing temperature suggesting less band admixture.
}
\end{figure*}
%

\section{Polarization dependent photoluminescence emission}

%
\begin{figure*}[ht]
\scalebox{\figurescale}{\includegraphics[width=1\linewidth]{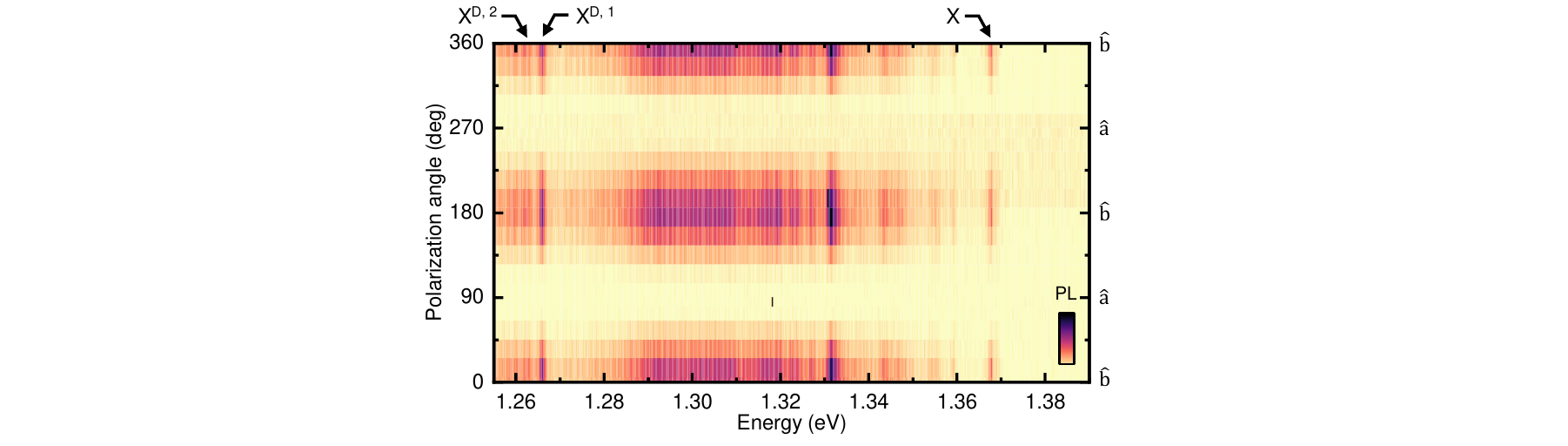}}
\renewcommand{\figurename}{SI Fig.|}
\caption{\label{SIfigPol}
\textbf{Polarization dependent PL emission of bulk CrSBr.} 
False color plot of the polarization angle-dependent low temperature ($T = \SI{4.2}{\kelvin}$) PL of bulk CrSBr. The exciton $X$ and the defect emission $X^{D,1}$ and $X^{D,2}$ are highlighted. All emission follows the $b$ axis which corresponds to the dispersive $Y$ direction in momentum space.~\cite{Klein.2022}
}
\end{figure*}
%

\newpage

\section{Real-space electronic defect wavefunction}

%
\begin{figure*}[ht]
\scalebox{\figurescale}{\includegraphics[width=0.6\linewidth]{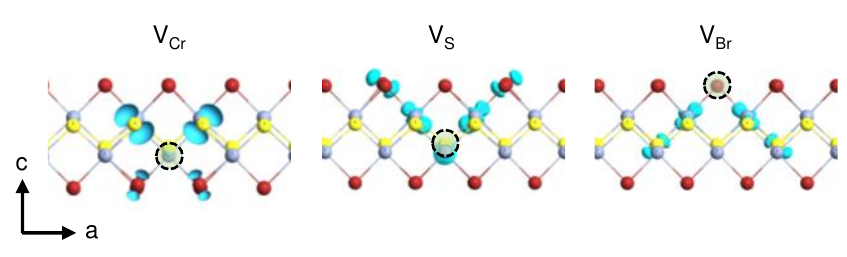}}
\renewcommand{\figurename}{SI Fig.|}
\caption{\label{SIfigwavefunctions}
\textbf{Charge density of the most prevalent vacancy defects in CrSBr.} 
Calculated electronic charge density of different defect states in real-space of the V\textsubscript{Cr}, V\textsubscript{S} and V\textsubscript{Br}.
}
\end{figure*}
%

\section{Anisotropy of the electronic structure of defect states}

%
\begin{figure*}[ht]
\scalebox{\figurescale}{\includegraphics[width=1\linewidth]{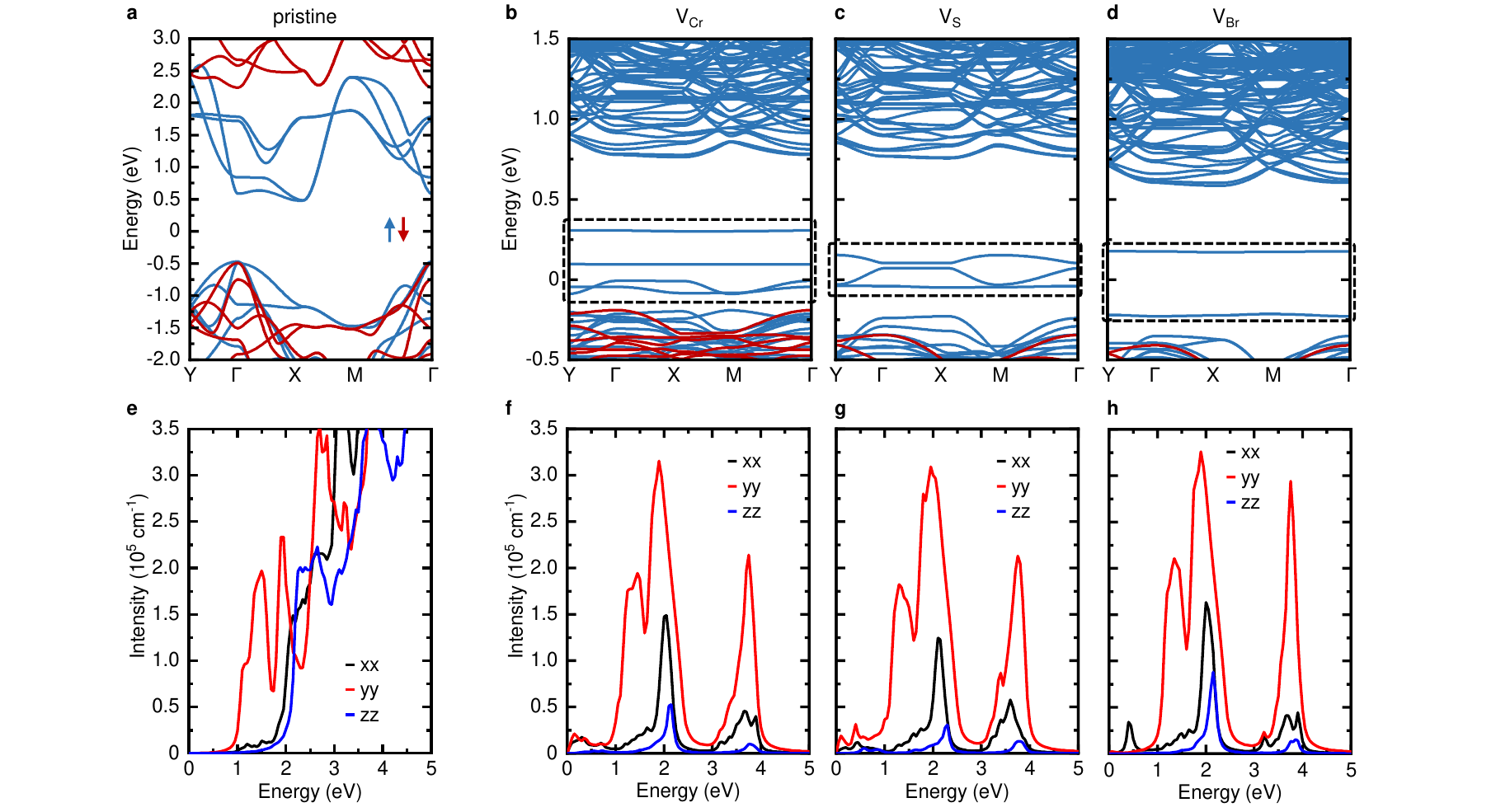}}
\renewcommand{\figurename}{SI Fig.|}
\caption{\label{SIfigabsorption}
\textbf{Anisotropy in the absorption from the electronic structure of the most prevalent point vacancy defects.} 
\textbf{a}, DFT-PBE electronic bandstructure of pristine monolayer CrSBr for a 7 $\times$ 7 $\times$ 1 supercell. The majority and minority spin of the bands is shown in blue and red colors.
\textbf{b}, Electronic bandstructure of V\textsubscript{Cr}, \textbf{c}, V\textsubscript{S} and \textbf{d} V\textsubscript{Br}. The defect levels are highlighted.
\textbf{e-h}, Corresponding calculated absorption for three different crystallographic orientations. Here, $xx$ corresponds to the $a$ axis, $yy$ to the $b$ axis and $zz$ to the $c$ axis. The absorption for the pristine CrSBr close to the band edge shows the expected strong polarization along the $b$ axis. The V\textsubscript{Cr} shows the least polarization while the V\textsubscript{S} is predominantly polarized along $b$ axis and the V\textsubscript{Br} along the $a$ axis. The data mostly resolves the defect-to-defect transitions while the defect-to-valence band or conduction band-to-defect transition are not straightforward to distinguish energetically from band-to-band transitions. The absorption data in combination with the strong extent of the electronic wavefunction along the $b$ axis and the experimentally determined polarization of the defect emission along the $b$ axis suggests the V\textsubscript{S} as the most likely candidate for the optical defect emission.
}
\end{figure*}
%
\newpage

\section{Orbital admixture of point vacancy defect states}

%
\begin{figure*}[ht]
\scalebox{\figurescale}{\includegraphics[width=1\linewidth]{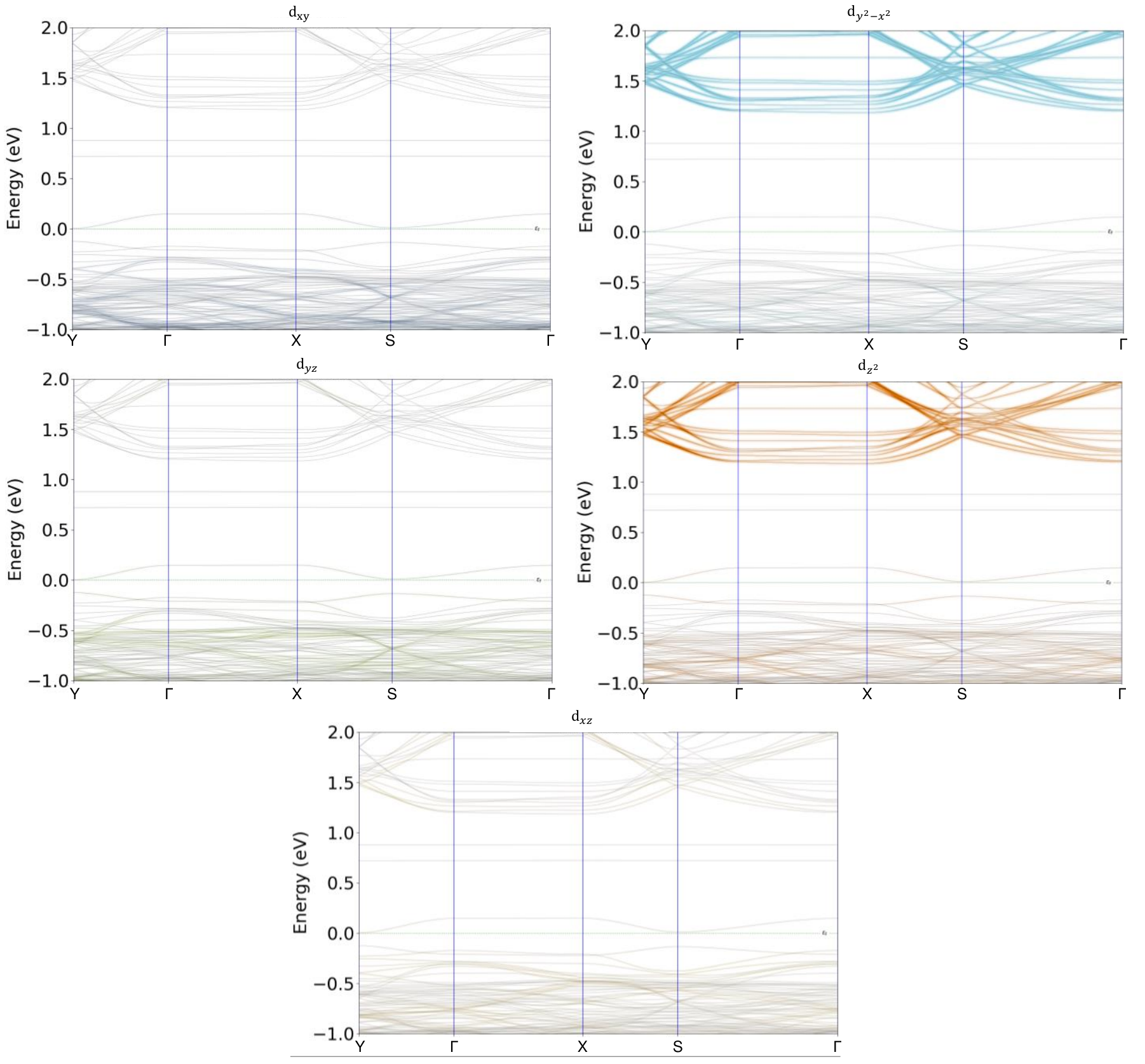}}
\renewcommand{\figurename}{SI Fig.|}
\caption{\label{SIfigCr}
\textbf{Admixture of chromium $d$-orbitals of the chromium vacancy defect V\textsubscript{Cr}.} 
Calculated electronic band structure of the V\textsubscript{Cr} with the orbital projection from the Cr $d$-orbitals: $d_{xy}$, $d_{y^2 - x^2}$, $d_{yz}$, $d_{z^2}$ and $d_{xz}$. The $d$-orbital admixture into the four defect bands is weak due to the strong localization of the electronic wavefunction of V\textsubscript{Cr}. For the calculation a 7 $\times$ 7 $\times$ 1 supercell was used. 
}
\end{figure*}
%

\newpage

%
\begin{figure*}[ht]
\scalebox{\figurescale}{\includegraphics[width=1\linewidth]{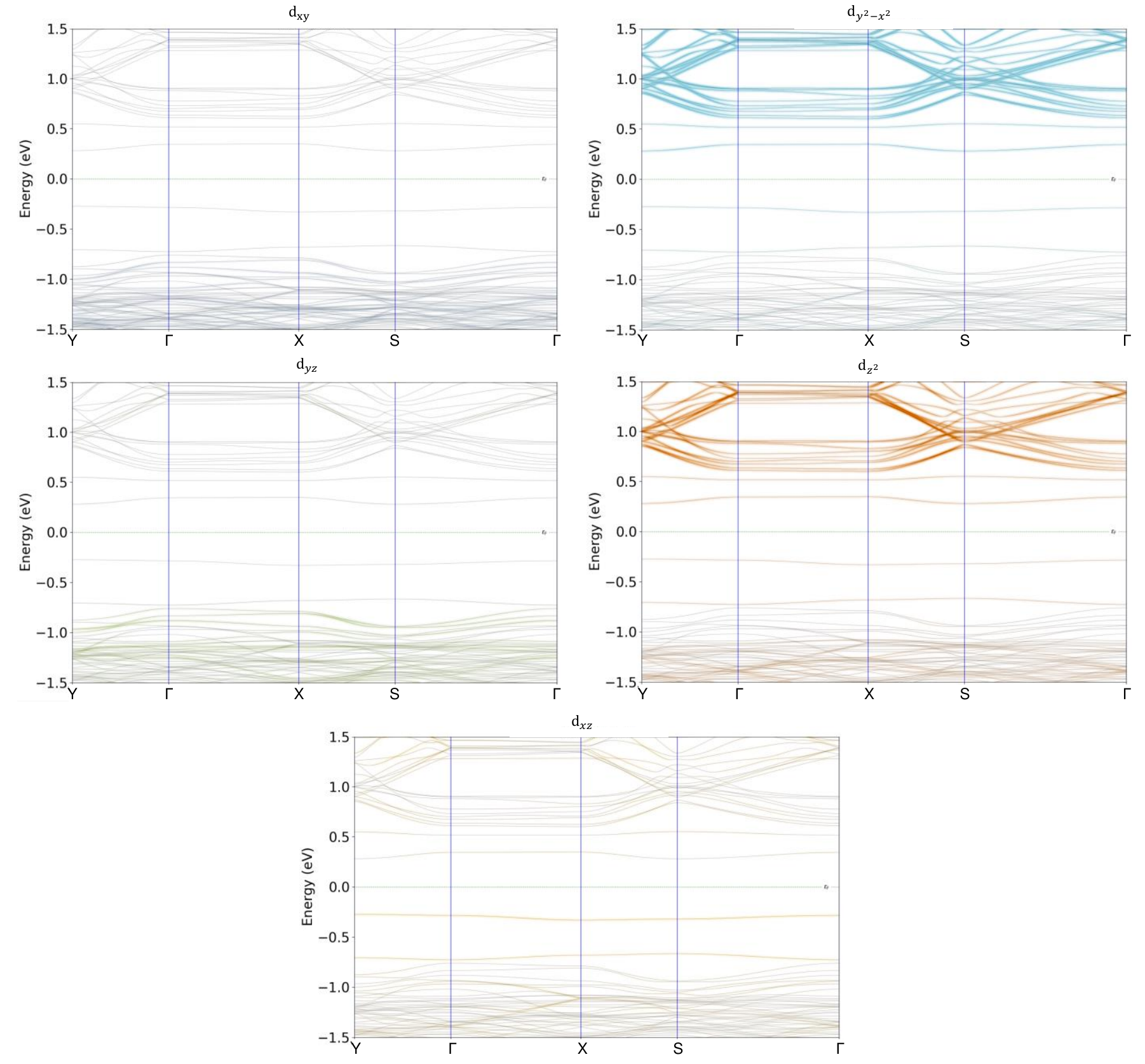}}
\renewcommand{\figurename}{SI Fig.|}
\caption{\label{SIfigS}
\textbf{Admixture of chromium $d$-orbitals of the sulfur vacancy defect V\textsubscript{S}.} 
Calculated electronic band structure of the V\textsubscript{S} with the orbital projection from the Cr $d$-orbitals. The predominant admixture to the four defect bands is from the $d_{y^2 - x^2}$, $d_{z^2}$ and $d_{xz}$. In particular, the two upper defect bands that are situated close to the conduction band exhibit predominant admixture from $d_{y^2 - x^2}$- and $d_{z^2}$-orbitals. In contrast, the lower two bands exhibit predominant admixture from the $d_{xz}$-orbitals.
}
\end{figure*}
%

\newpage

%
\begin{figure*}[ht]
\scalebox{\figurescale}{\includegraphics[width=1\linewidth]{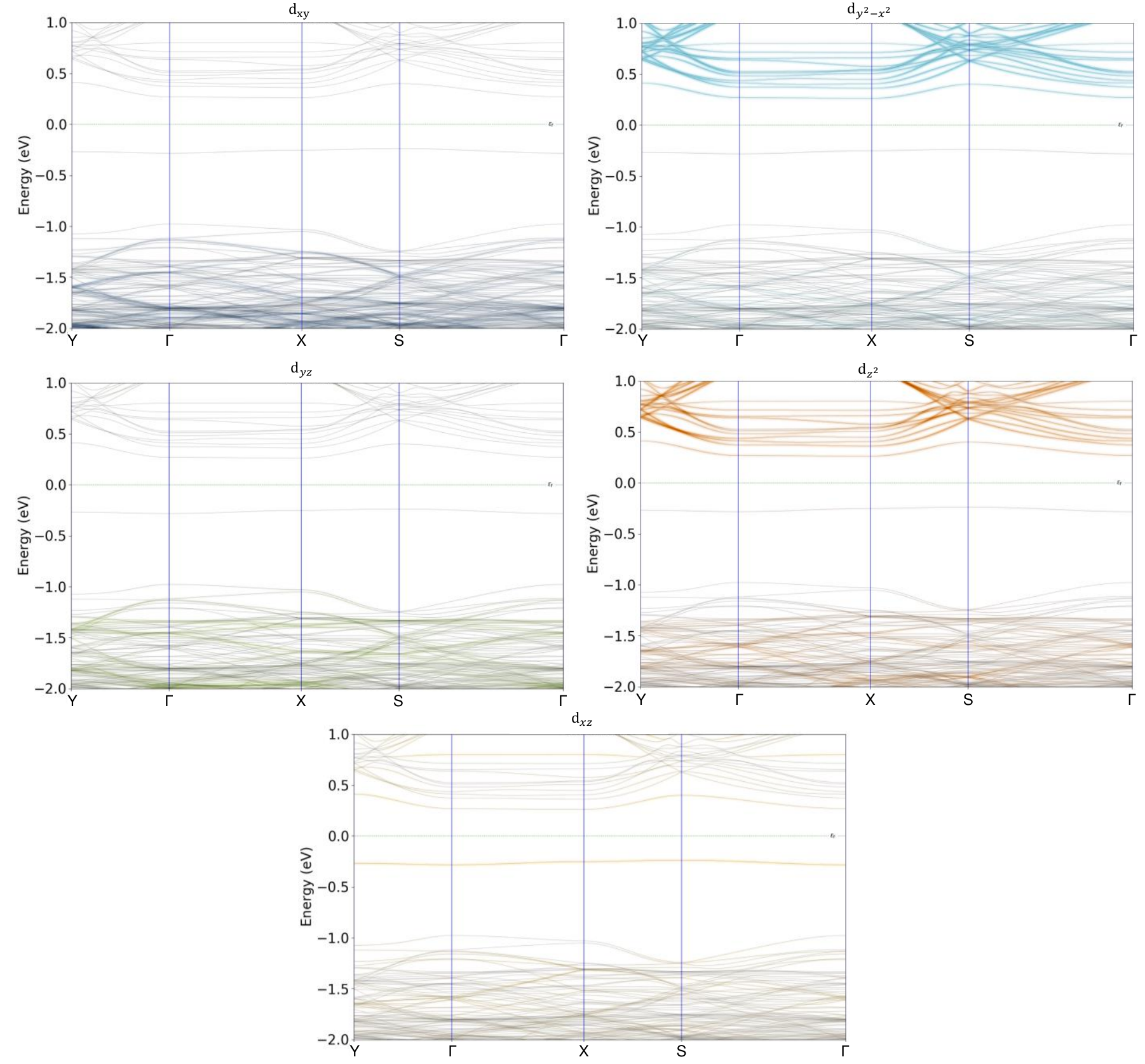}}
\renewcommand{\figurename}{SI Fig.|}
\caption{\label{SIfigBr}
\textbf{Admixture of chromium $d$-orbitals of the bromine vacancy defect V\textsubscript{Br}.} 
Calculated electronic band structure of the V\textsubscript{Br} with the orbital projection from the Cr $d$-orbitals. The V\textsubscript{Br} exhibits two midgap bands. The predominant admixture to the two defect bands is from the $d_{y^2 - x^2}$, $d_{z^2}$ and $d_{xz}$ similar to the V\textsubscript{S}. In particular, the upper defect band is situated close to the conduction band and exhibits predominant admixture from $d_{y^2 - x^2}$- and $d_{z^2}$-orbitals. In contrast, the lower band exhibits predominant admixture from the $d_{xz}$-orbitals.
}
\end{figure*}
%

\newpage

\section{Theory for Defect-Carrier Interactions}

Here we explore a possible origin of the magnetic interactions between defects.
We assume that the defect energy is sufficiently close to the chemical potential that the defect orbitals are occupied at least partially.
The occupied defects are then expected to host an embedded spin which we model classically here.
Through an $s-d$ exchange interaction with the conduction band (or the valence band, which behaves similarly for our purposes), these spins can couple to itinerant carriers, which we model via a contact exchange interaction with the spin density. 

Since the carriers have a very large anisotropy in effective mass,~\cite{Klein.2022} we consider a simple model of one-dimensional carriers with effective mass $m^* = 0.14 m_e$ interacting with defects randomly located at points $\pm R/2$.
We model the simplest case here, and leave further details to future work. 
First, we consider the case of a lone defect at $x=0$ interacting with the conduction band, with one-body carrier Hamiltonian 
\begin{equation}
    \hat{H}_{\rm c}[S_1] = -\frac{\hbar^2}{2m^*}\frac{\partial^2}{\partial x^2}+\frac12 \Delta (\hat{\sigma}_z+1) + \frac{1}{2} {\vec \sigma}\cdot\mathbf{S} K_{\rm d} \delta(x).
\end{equation}
Here ${\vec \sigma}$ is the spin operator for the carrier electron, $\Delta$ is the exchange splitting of the conduction band, $K_{\rm d}$ is the defect-carrier $s-d$ exchange interaction, and $\mathbf{S}$ is the expectation value of the defect spin, which we treat classically. 
We can convert the exchange interaction into the scattering length parameter $K_{\rm d} = \frac{1}{m^*a}$. 
We take $K_{\rm d} > 0$, since the opposite sign is equivalent to a global inversion of all the impurity spins, which can be done without inducing frustration. 

We assume the exchange splitting is very large, such that we can project onto the majority spin state, which then reduces the Hamiltonian to 
\begin{equation}
    \hat{H}[S_1] = -\frac{\hbar^2}{2m^*}\frac{\partial^2}{\partial x^2}+ \frac{1}{2}S^z K_{\rm d} \delta(x).
\end{equation}
If $S^z$ is anti-aligned with the carrier spin axis, then this leads to an attractive delta-function potential while it is repulsive if it is aligned.
Formally, we can solve for the wavefunctions of this potential by writing 
\begin{equation}
    \psi(x) = \begin{cases} 
    A e^{\kappa x} & x < 0 \\
    A e^{-\kappa x} & x > 0. \\
    \end{cases}
\end{equation}
This is solved by imposing continuity of the wavefunction at 0, as well as setting the discontinuity of the derivative. 
We find 
\begin{equation}
    \frac{\kappa}{m^*} + \frac{S^z}{2m^* a} = 0 
\end{equation}
and the requirement that $\kappa > 0$ means this only has solutions if $S^z < 0 $, corresponding to the anti-aligned configuration. 
This then gives, for impurity spin length $S$,
\begin{equation}
    \kappa = - \frac{S}{2a}.
\end{equation}
The energy of this eigenstate is found in terms of $\kappa $ to be 
\begin{equation}
    E_{\downarrow} = -\frac{\hbar^2 \kappa^2}{2m^*} = -\frac{\hbar^2}{8m^* a^2}S^2.
\end{equation}
Thus, we find that there is essentially a Zeeman-induced spin-splitting from the polarized carrier band of 
\begin{equation}
    E_{\uparrow} - E_{\downarrow} = \frac{\hbar^2}{8m^* a^2}S^2. 
\end{equation}

Now, we consider the case of two defects.
We take the same parameters, and solve for the bound states in a similar way.
The one-body carrier Hamiltonian is (again projected on to the majority spin species)
\begin{equation}
    \hat{H}[S_1,S_2] = -\frac{\hbar^2}{2m^*}\frac{\partial^2}{\partial x^2}+ \frac{1}{2m^* a}S^z_1 \delta(x+R/2) + \frac{1}{2m^* a} S^z_2\delta(x - R/2).
\end{equation}
We now write for the wavefunction
\begin{equation}
    \psi(x) = \begin{cases} 
    A e^{\kappa x} & x <-R/2 \\
    B_+ e^{\kappa x} + B_{-}e^{-\kappa x} & x \in [-R/2,R/2] \\
    C e^{-\kappa x} & x > R/2. \\
    \end{cases}
\end{equation}
The equations for continuity of the wavefunction imply 
\begin{subequations}
\begin{align}
    & A e^{-\kappa R/2} = B_+ e^{-\kappa R/2} + B_{-} e^{\kappa R/2} \\
    & C e^{-\kappa R/2} = B_+ e^{\kappa R/2} + B_{-} e^{-\kappa R/2} .
\end{align}
\end{subequations}
We next impose the derivative matching conditions of
\begin{subequations}
\begin{align}
    & -\frac{1}{2m^*}(\kappa B_+ e^{-\kappa R/2} -\kappa B_-e^{\kappa R/2} - \kappa Ae^{-\kappa R/2})  + \frac{1}{2m^* a}S^z_1 Ae^{-\kappa R/2} =0  \\
    & -\frac{1}{2m^*}(-\kappa C e^{-\kappa R/2} - \kappa B_+ e^{\kappa R/2}+\kappa B_-e^{-\kappa R/2} )  + \frac{1}{2m^* a}S^z_2 Ce^{-\kappa R/2} =0 .
\end{align}
\end{subequations}
Together this is a system of four equations for four coefficients, and in general this imposes a secular equation on the eigenvalue $\kappa$.
After algebra we find 
\begin{equation}
    \left( \coth\kappa R + 1 +\frac{S^z_1}{\kappa a} \right)\left( \coth \kappa R + 1 + \frac{S^z_2}{\kappa a}\right) - \frac{1}{\sinh^2\kappa R} = 0.
\end{equation}
The bound state energy is one again found to be 
\begin{equation}
    E[S_1,S_2] = -\frac{\hbar^2 \kappa^2}{2m^*}
\end{equation}
where $\kappa$ is found by solving the above equation in terms of the $S^z_1,S^z_2$. 
We can then calculate the energy of the different spin configurations as well as the wavefunctions. 
For the case of antialigned defect spins $S^z_1 = +1,S^z_2 = -1$, we find the bound state is essentially connected to the single-defect bound state located on $S^z_2$, as clearly seen by taking the separation to be large, in which case the bound state energy essentially recovers the single-defect energy.
On the other hand, the bound state for the two aligned defects has an energy which is lower than the single-defect energy even up to large separations, supporting the idea that when the defects are spin aligned they allow for a greater delocalization of the bound states and thus are energetically favorable beyond the single-defect binding energy.
This would support the idea that defect spins actually interact through exchange of their bound magnetic polarons, and exhibit a genuine phase transition~\cite{Liu.1986,Kaminski.2002,Durst.2002,Calderon.2007,Tang.2007,Priour.2006,Bednarski.2012}.
However, in order to investigate this more thoroughly we need to consider a more complicated model which accounts for the role of non-contact interactions, multiple carriers, and carrier interactions.
It is also still to be seen whether this model can  explain quantitatively the strongly asymmetric broadenings of the defect lines.
This is an interesting problem which we leave to future studies.

%
\newpage

%
%
%
%
\bibliographystyle{apsrev}
\bibliography{full}